\documentclass{article}

\usepackage[utf8]{inputenc}
\usepackage[margin=1in]{geometry}
\usepackage[titletoc,title]{appendix}

\usepackage{amsmath,amsfonts,amssymb,mathtools}

\usepackage{graphicx,float}

\usepackage{multirow}
\usepackage{array}

\usepackage[ruled,vlined]{algorithm2e}
\usepackage{algorithmic}

\usepackage{cite}

\title{Data-driven modeling of an oscillating surge wave energy converter using dynamic mode decomposition}

\author{$\text{Brittany Lydon}^{*1}$, $\text{Brian Polagye}^{1}$, $\text{Steven Brunton}^{1}$}
\date{June 6th, 2023}

\begin{document}

\maketitle

\begin{centering}
$^*$Corresponding author email: BrittLyd@uw.edu \\
Contributing authors email: bpolagye@uw.edu, sbrunton@uw.edu \\
$^1$ University of Washington, Department of Mechanical Engineering, Seattle, WA, USA \\
\end{centering}

\begin{abstract}
    Modeling wave energy converters (WECs) to accurately predict their hydrodynamic behavior has been a challenge for the wave energy field. Often, this results in either low-fidelity, linear models that break down in energetic seas, or high-fidelity numerical models that are too computationally expensive for operational use. To bridge this gap, we propose the use of dynamic mode decomposition (DMD) as a purely data-driven technique that generates an accurate and computationally efficient model of an oscillating surge WEC (OSWEC). Our goal is to model and predict the behavior of the OSWEC in monochromatic and polychromatic seas without knowledge of the governing equations or incident wave field. We generate the data for the algorithm using a semi-analytical model and the open-source code WEC-Sim, then evaluate how well DMD can describe past dynamics and predict future state behavior. We consider realistic challenges including noisy sensor measurements, nonlinear WEC dynamics, and irregular wave forcing. In each of these cases, we generate accurate models for past and future OSWEC behavior using DMD, even with limited sensor measurements. These findings provide insight into the use of DMD on systems with limited time-resolved data and present a framework for applying similar analysis to lab- or field-scale experiments.
\end{abstract}

\section{Introduction}
\label{sec:introduction}

\begin{figure}[t!]
    \centering
    \includegraphics[width=1\linewidth]{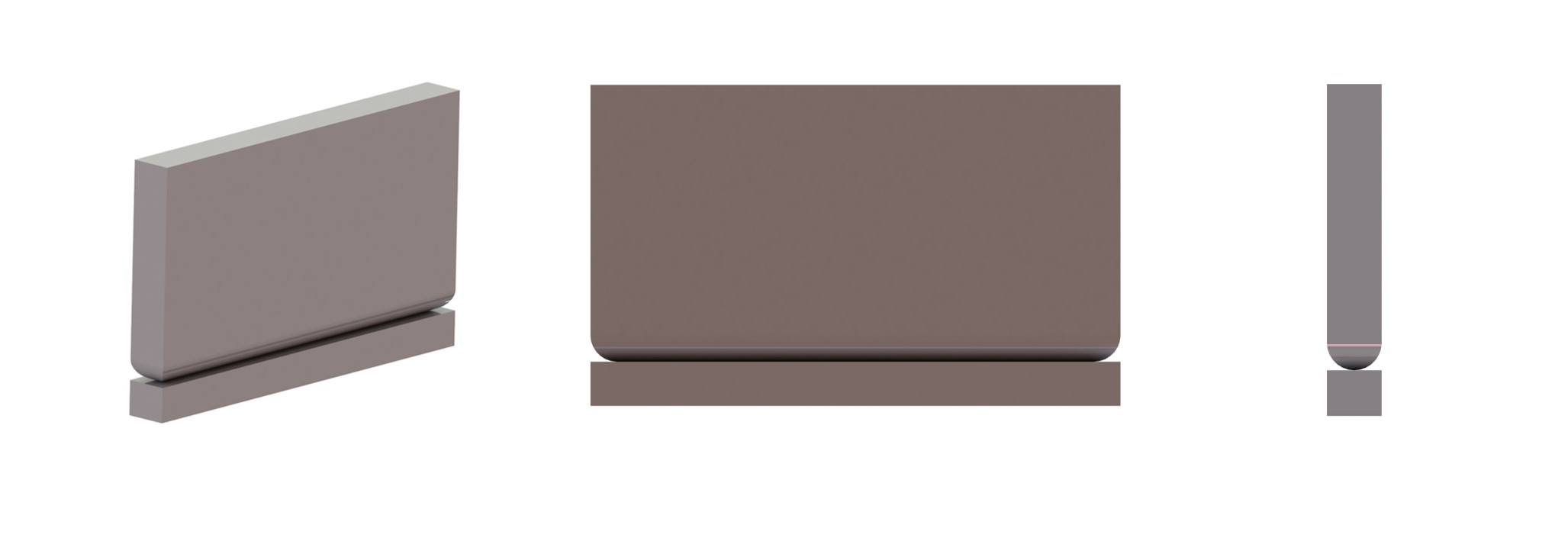}
    \caption{Isometric (left), front (middle), and side (right) view of a generic OSWEC device. OSWECs are made up of two bodies, a flap and base, and the flap oscillates in pitch about a hinge located between the two bodies.}
    \label{fig:oswec}
\end{figure}

Wave energy converters (WECs) have promise for carbon-free energy generation, especially for coastal communities~\cite{manasseh2017integration}. Although there has been significant innovation in the wave energy field, modeling and controlling WEC behavior continues to be a difficult task due to the complicated fluid-structure interaction between the device and the waves and the volatile nature of the ocean~\cite{aderinto2018ocean}. A promising WEC technology is the oscillating surge wave energy converter (OSWEC)~\cite{folley2004oscillating} due to its ability to absorb power over a wide range of wave frequencies~\cite{folley2015contrasting}. An example of a generic OSWEC is shown in Figure \ref{fig:oswec}. OSWECs are flap-type devices that primarily harness the surge motion of ocean waves, which causes a buoyant flap to oscillate in pitch about a hinge~\cite{whittaker2007development}. OSWECs are particularly promising in shallow ocean environments, where the wave orbitals elongate due to shoaling, amplifying surge motion, and the wave direction is consistent~\cite{falnes2005ocean}. OSWECs are also particularly attractive devices for nearshore desalination, as the mechanical pitching motion can be directly translated to a pressure differential for reverse osmosis without losses from intermediate electricity generation~\cite{folley2009cost}. 

Despite the benefits of OSWECs, there are significant challenges to their development, as evidenced by the commercial failure of the Aquamarine Oyster in 2015~\cite{whittaker2007development}. Although OSWECs operate in a single degree of freedom, the dynamics and resulting kinematics are difficult to describe in realistic sea states. This is partly because diffraction is one of the dominant forces that drives OSWEC dynamics, making common assumptions such as small-body approximations invalid for these systems~\cite{giorgi2016nonlinear}. The dynamics are also influenced by drag and viscous forces, resulting in complex fluid-structure interaction that can lead to errors in modeling and predicting energy absorption~\cite{babarit2012numerical}. Because of these complexities, common design and modeling practices are often not suitable for OSWECs~\cite{folley2015contrasting}. For example, models based in linear wave and potential flow theory break down quickly with OSWECs due to the underlying mechanisms driving their hydrodynamics~\cite{folley2015contrasting}. Because of this, there is a significant body of work that focuses on developing high-fidelity models using computational fluid dynamics (CFD)~\cite{wei2015numerical,wei2016wave,schmitt2015use} and smoothed particle hydrodynamics (SPH)~\cite{wei2013numerical,wei2015numerical,chen2014numerical,yeylaghi2016isph,zhang2018sph,wei2019modeling,brito2020numerical} to more accurately predict OSWEC behavior. However, these models can take hours of clock time to simulate a single oscillation period of WEC behavior, making them unsuitable for real-time state estimation and control schemes~\cite{brunton2022data}. This motivates modeling techniques for OSWECs that are both accurate enough to model these systems in realistic waves, and fast enough to be used in a real-time control scheme, such as model predictive control (MPC)~\cite{Kaiser2018prsa}. 

Data-driven modeling techniques are well suited to fill this role, particularly dynamic mode decomposition (DMD)~\cite{schmid2010dynamic,rowley2009spectral,tu2013dynamic,kutz2016dynamic}. DMD is an equation-free modal decomposition technique that can model and predict the behavior of complex dynamical systems, including fluid flows~\cite{schmid2010dynamic, schmid2011application, mezic2013analysis}, as well as create robust and computationally efficient models for optimal control~\cite{proctor2016dynamic, korda2018linear, sharan2022real}. DMD is an established and powerful method in the fields of robotics~\cite{berger2015estimation}, video processing~\cite{grosek2014dynamic, kutz2016video, erichson2015compressed}, and neuroscience~\cite{brunton2016extracting}, and has been effective in modeling experimental systems~\cite{schmid2011application, schmid2010dynamic}. 

There has been some recent work to utilize DMD for ocean engineering problems. For example, Serani et al.~\cite{serani2023use} and Diez et al.~\cite{diez2022time} successfully applied DMD for ship maneuvering in both regular and irregular waves. Both studies found that models generated from DMD were able to accurately forecast ship forces and trajectories for various ship operations in multiple wave periods. In general, the studies found while DMD works quite well for situations with predominantly monochromatic dynamics, irregular wave inputs and nonlinear dynamics can reduce the accuracy of the DMD forecasts and often benefit from variants of DMD. This suggests care must be taken when considering realistic ocean behaviors to preserve forecasting accuracy. DMD has not, to the authors' knowledge, been directly applied to wave energy conversion, but this previous work shows promise in addressing barriers to WEC modeling and control.  

In its most general form, DMD decomposes spatial-temporal time series data into spatially coherent dynamic modes with associated complex eigenvalues that determine the frequency and growth/decay rates of the respective mode behavior in time. These modes and eigenvalues can be used to build reduced-order models of dynamical systems and forecast system behavior, all without knowing the governing equations of the system. DMD eigenvalues and eigenvectors approximate those of the Koopman operator~\cite{koopman1931hamiltonian,Mezic2005nd,rowley2009spectral,mezic2013analysis,Brunton2017natcomm,Brunton2022siamreview,colbrook2022mpedmd,colbrook2023residual}, which is an infinite-dimensional linear operator that describes the dynamics of a fully nonlinear system. DMD contains properties of both proper orthogonal decomposition (POD) and the Fourier transform, as it identifies spatial modes (characteristic of POD) whose time behavior is characterized by a corresponding frequency (characteristic of the Fourier transform)~\cite{chen2012variants, proctor2016dynamic}. DMD has three major applications: identifying and characterizing system behavior, state estimation and/or future state prediction, and control~\cite{kutz2016dynamic}. 

DMD is well suited to model WEC hydrodynamics for three main reasons:

\begin{itemize}
\item \textbf{Equation-free modeling.} DMD generates a system model purely from time-resolved data, which means we can model and predict WEC behavior without knowing or solving the underlying governing equations. Since OSWEC dynamics are not always well described with classical analytical techniques~\cite{folley2015contrasting}, DMD can identify dynamics that are not well captured by linear simplifications of OSWEC governing equations.

\item \textbf{Simple, but adaptable.} DMD is based in straight-forward linear algebra, resulting in low computational cost, as well as interpretable outputs. DMD also has extensions that can address common issues in practical applications, such as noisy signals~\cite{bagheri2014effects,hemati2017biasing, dawson2016characterizing,askham2018variable} (further discussed in Section \ref{sec:TLSDMD}), limited sensor availability~\cite{brunton2013compressive, tu2014spectral}, and evolution over multiple time scales~\cite{kutz2016multiresolution}, all of which are present for OSWECs.

\item \textbf{Optimal control.}  Finally, because DMD can forecast system behavior, the reduced-order model generated from DMD can be used in a model predictive control scheme~\cite{proctor2016dynamic,Kaiser2018prsa}. This is a particularly attractive option for WECs since optimal control schemes have shown to increase the efficiency of different WEC archetypes compared to passive control methods~\cite{andersen2015model, richter2014power} and traditional linear models may not be accurate or fast enough to be used for this application~\cite{bacelli2014nonlinear, merigaud2017optimal}. 
\end{itemize}

Our objective is to generate a data-driven model using DMD that can predict OSWEC behavior without incurring an infeasible computational cost or sacrificing accuracy. In particular, we aim for this model to address well known issues in WEC modeling, including handling noisy signals, modeling nonlinear WEC dynamics, and modeling WEC behavior in response to irregular seas. The remainder of the paper is laid out as follows: Section \ref{sec:methods} describes the methods we used in this study, as well as the three main cases we are testing. Section \ref{sec:results} presents the results and discussions of DMD performance on OSWEC behavior, with an emphasis on treatment of realistic challenges. Specifically, Section \ref{sec:noise} considers DMD with noisy data, Section \ref{sec:nonlinear} considers DMD when weakly nonlinear dynamics are present, and Section \ref{sec:irregular} considers DMD modeling of an OSWEC in an irregular wave field.

\section{Methods}
\label{sec:methods}

\begin{figure}[t!]
    \centering
    \includegraphics[width=1\linewidth]{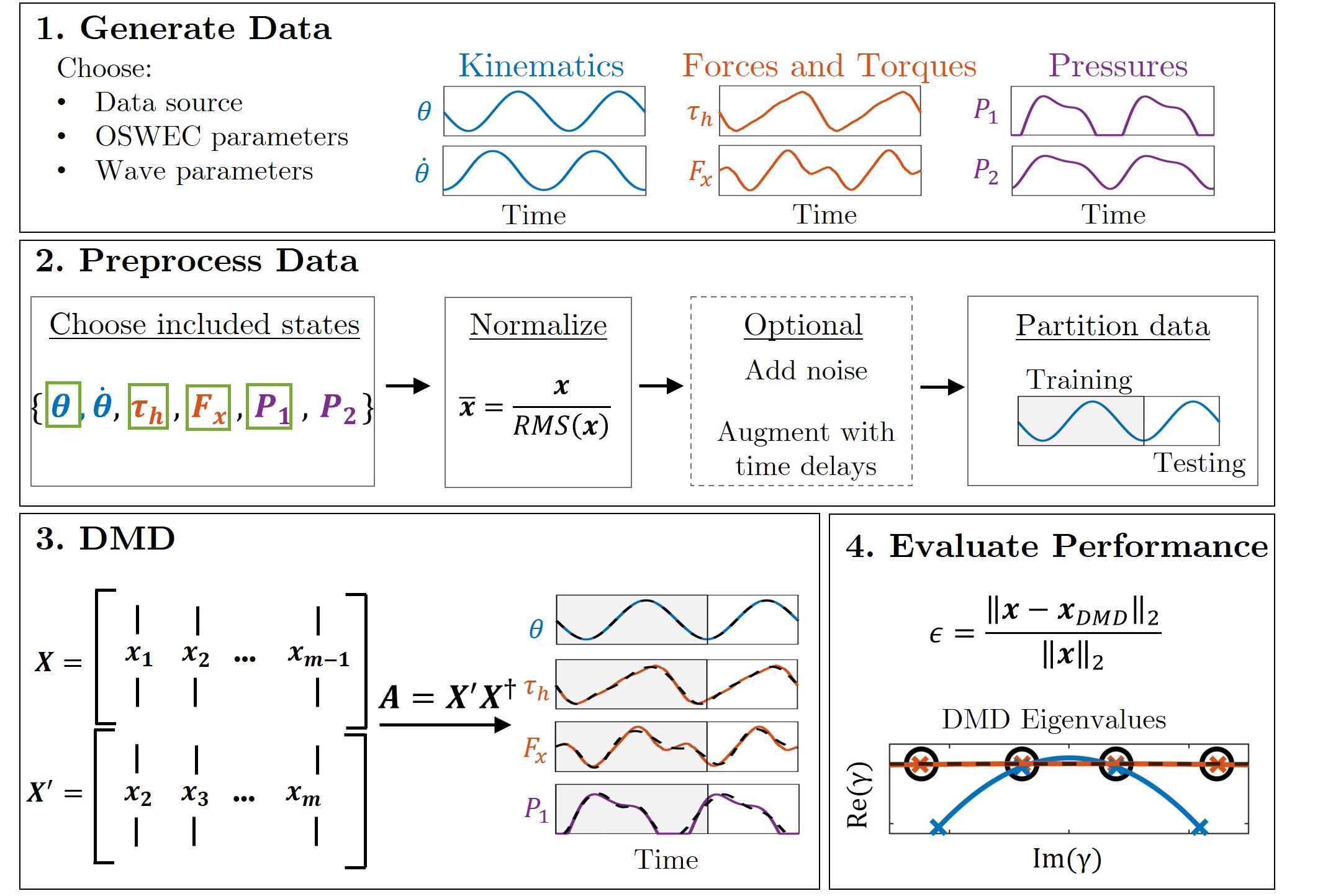}
    \caption{Data analysis workflow. The first step is to generate data by choosing the data source (analytical model, CFD, etc.) and wave input. Next, we preprocess data to prepare for DMD algorithm. We then run DMD on training data, and finally we evaluate the performance of the algorithm on both the training and testing data using a nondimensionalized error parameter, $\epsilon$.}
    \label{fig:workflow}
\end{figure}

An overview of the methods we used for this study are outlined in Figure \ref{fig:workflow}. There are four major steps, each outlined in detail in the following subsections: generate data, preprocess data, run DMD, and evaluate performance. 

\subsection{Generate data}
The first step of our workflow is to generate the data that will be used to train the DMD model (Block 1 in Figure \ref{fig:workflow}). This step includes choosing the data source, which OSWEC parameters to include, and wave conditions. We use two synthetic data sources in this study: a semi-analytical model developed by Renzi et al.~\cite{renzi2013hydrodynamics} and a mid-fidelity modeling tool, WEC-Sim~\cite{wecsim} (further details in Section \ref{sec:sam} and \ref{sec:wecsim}, respectively). Although we are using low- and mid-fidelity models to generate OSWEC dynamics, this same process can be used with limited variations for high-fidelity CFD models or experimental data.

\begin{figure}[t!]
    \centering
    \includegraphics[width=1\linewidth]{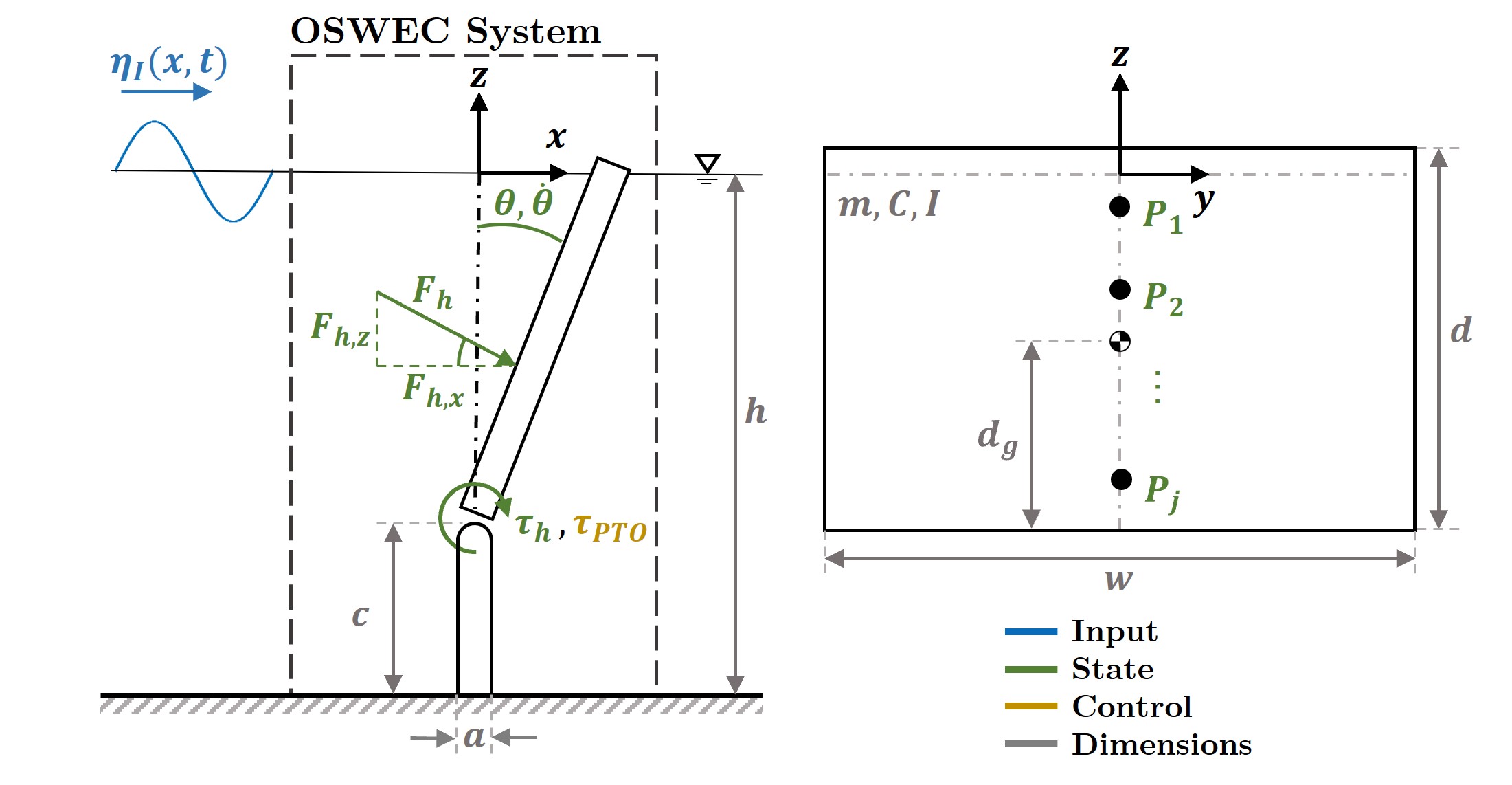}
    \caption{Diagram of OSWEC system with incident wave input in blue, system states in green, control parameters in yellow, and dimensions in gray.}
    \label{fig:oswec_system}
\end{figure}

A diagram of the OSWEC system we use in this study is shown in Figure \ref{fig:oswec_system} and consists of a surface-piercing flap oscillating in the x-z plane with input (shown in blue), system states (shown in green), and control input (shown in yellow). For all of our data generation, we consider a utility-scale OSWEC analogous to the Aquamarine Oyster~\cite{whittaker2007development} with a flap width of 18 m operating in a still water depth of 10.9 m. Relevant OSWEC system dimensions and mass properties shown in gray in Figure \ref{fig:oswec_system} are enumerated in Table \ref{tab:oswec_parameters}, including moment of inertia ($I$) and hydrostatic stiffness ($C$) of the flap. 

\begin{table}[]
    \centering
    \caption{OSWEC system dimensions.}
    \begin{tabular}{|m{4.5cm}|m{2.5cm}|}
        \hline
        \hfil $\mathbf{Parameter}$ & \hfil $\mathbf{Value}$ \\
        \hline
        \hfil still water depth, $h$ & \hfil 10.9 m \\
        \hline
        \hfil flap width, $w$ & \hfil 18 m \\
        \hline
        \hfil flap height, $d$ & \hfil 9.4 m\\
        \hline
        \hfil flap thickness, $a$ & \hfil 1.8 m \\
        \hline
        \hfil foundation height, $c$ & \hfil 1.5 m \\
        \hline
        \hfil CoG to hinge distance, $d_{g}$ & \hfil 5 m \\
        \hline
        \hfil flap mass, $m_{f}$ & \hfil 127 * $10^{3}$ kg \\
        \hline
        \hfil flap moment of inertia, $I$ & \hfil 1.85 * $10^{6}$ kg m$^{2}$ \\ 
        \hline
        \hfil flap hydrostatic stiffness, $C$ & \hfil 6.4 * $10^{6}$ N m \\
        \hline
    \end{tabular}
    \label{tab:oswec_parameters}
\end{table}

The system states in Figure \ref{fig:oswec_system} are shown in green. These states represent time series measurements that describe system behavior and are known for the purposes of DMD (these form our data matrices in the DMD algorithm). We chose the states of this system to be quantities that can be realistically measured in OSWEC systems and have been measured in previous physical studies~\cite{henry2014two,schmitt2012hydrodynamic,lamont2015investigating}. The kinematic states include angular position, $\boldsymbol{\theta}$, and angular velocity, $\dot{\boldsymbol{\theta}}$. Other dynamic states include the hydrodynamic force acting on the flap (balance of excitation force with radiation force and buoyancy), $\boldsymbol{F}_{h}$, which we decompose in both the surge and heave direction, $\boldsymbol{F}_{x}$ and $\boldsymbol{F}_{z}$, respectively, reflecting a multi-axis load measurement in the foundation frame of reference. We also consider the hydrodynamic torque acting about the hinge, $\boldsymbol{\tau}_{h}$, as well as pressure, $\boldsymbol{P}_{j}$, from an array of sensors aligned with the midline of the flap in the cross-flow direction. 

The input to the system is the incident wave field, $\boldsymbol{\eta}_{I}(\boldsymbol{x},t)$, shown in blue in Figure \ref{fig:oswec_system}. This wave field can be regular or irregular and is a function of both space and time. For the purposes of DMD, the incident wave field is an unknown quantity. This reflects the practical difficulty of obtaining time-resolved wave field measurements, which is a general challenge in WEC modeling and control~\cite{faedo2017optimal}. Although we do not directly measure $\boldsymbol{\eta}_{I}(\boldsymbol{x},t)$, we know it has a direct effect on the system state parameters we are modeling, and therefore treat it as a hidden or latent variable in this system. 

The control parameter we are considering is the torque applied by the power-takeoff at the hinge, $\boldsymbol{\tau}_{PTO}$. For this study we consider a simple linear damping control scheme, where $\boldsymbol{\tau}_{PTO} = \nu_{PTO} \dot{\boldsymbol{\theta}}$, and $\nu_{PTO}$ is the linear damping coefficient. For all cases, we chose a linear damping coefficient of 12000 N m s. 

\subsubsection{Semi-analytical model}
\label{sec:sam}
The first modeling method we use to generate training data for the DMD algorithm is a semi-analytical model developed by Renzi et al. that describes linear OSWEC behavior in the open ocean~\cite{renzi2013hydrodynamics}. We chose this semi-analytical model because it represents the dynamics of the flap and wave interaction without the abstraction of a boundary element method (BEM) to describe the hydrodynamics. This model is based on linear potential flow theory, and solves a collocation scheme for frequency-dependent hydrodynamic coefficients including added mass, $\mu$, radiation damping, $\nu$, and the complex excitation torque coefficient, $\hat{T}_{e}$. The flap kinematics are then governed by the equation of motion in the frequency domain:
\begin{equation}
    [-\omega^2(I+\mu) + C - i\omega(\nu+\nu_{PTO})]\hat{\Theta} = \hat{T}_{e},
\end{equation}
 where $\omega$ is the wave frequency, $I$ is the flap moment of inertia, $C$ is the flap buoyancy stiffness, and $\hat{\Theta}$ is the complex coefficient of angular rotation. We can then solve for $\hat{\Theta}$ in the frequency domain, and transform each measurement back to the time domain to obtain the time series response of the OSWEC. Finally, we can estimate pressure along the flap using the time derivative of the flow potential at the flap interface, $P = -\rho {\partial \chi}/{\partial t}$, where $\chi$ is the flow potential and $\rho$ is the water density. 

\begin{figure}[t!]
    \centering
    \vspace{-.6in}
    \includegraphics[width=1\linewidth]{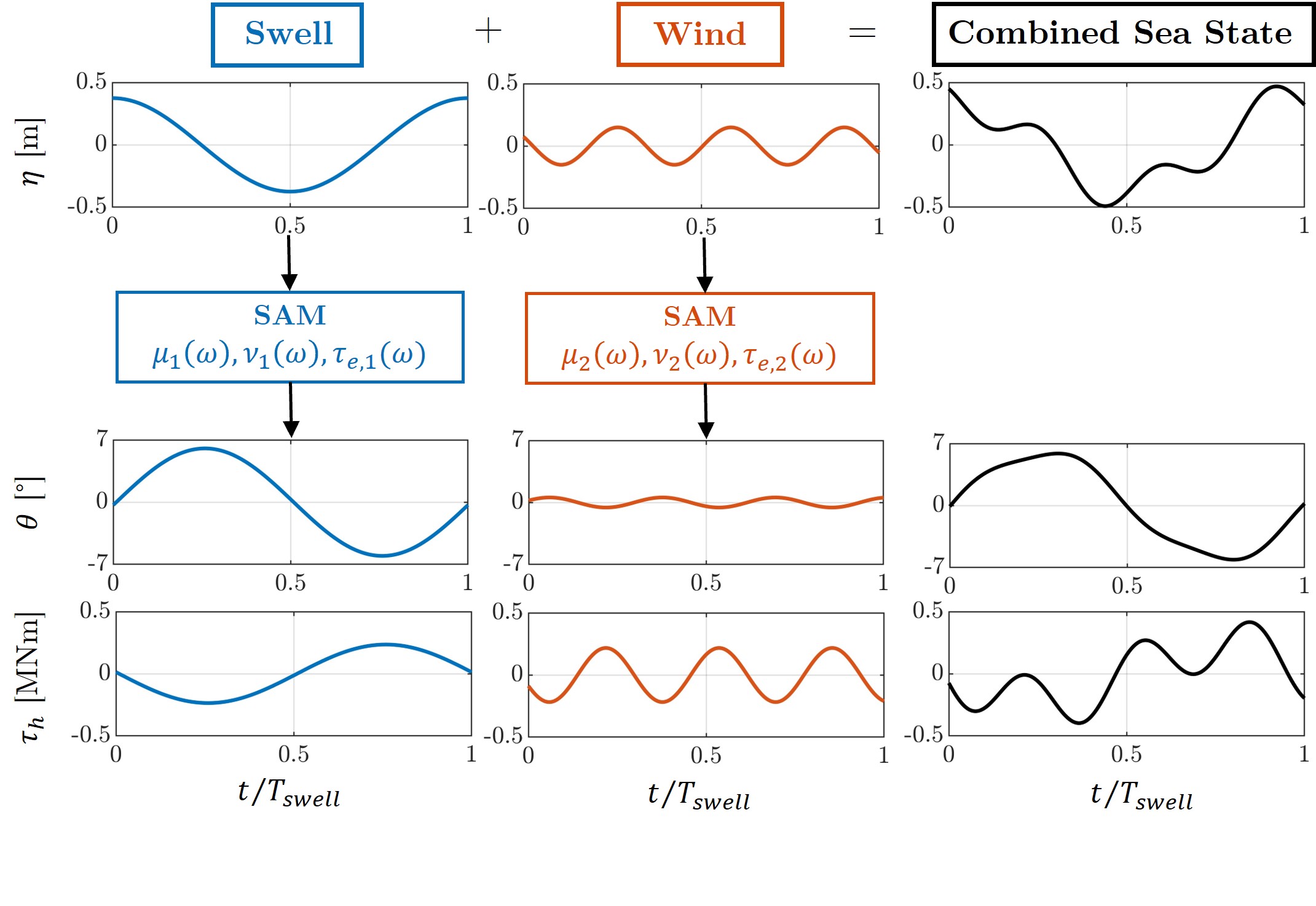}
    \vspace{-.6in}
    \caption{Process visualization for generating data with a polychromatic wave input from semi-analytical model (SAM)~\cite{renzi2013hydrodynamics} using superposition. The time duration is normalized by the period of the swell component, $T_{\mathrm{swell}}$. The top row represents the input into the SAM with the left column being the swell wave component in blue, and the middle column being the wind wave component in orange. The wave fields are then treated as individual inputs to the SAM. The middle and bottom rows show two output variables: angular rotation $\boldsymbol{\theta}$ and hydrodynamic torque $\boldsymbol{\tau}_{h}$, respectively. The far right column in black shows the superposition of the separate responses to each wave component.}
    \label{fig:dias_sum}
\end{figure}

This semi-analytical model was developed for a monochromatic wave input with single frequency, $\omega$ and wave height, $H$. To approximate a more realistic sea state, we adapted the model for a polychromatic wave input using linear superposition of two frequencies (Figure \ref{fig:dias_sum}). Specifically, we considered a polychromatic wave input composed of two monochromatic waves to model the combination of swell (longer period, larger wave height) and wind wave (shorter period, smaller wave height) behavior. Figure \ref{fig:dias_sum} shows these two wave components in the first row, where the swell wave component is in blue, the wind wave component in orange, and the combined state in black. Each wave component is used as an input to the semi-analytical model (labelled SAM), similar to sending an input signal through a transfer function. Once the outputs are transformed to the time domain, they are added together to approximate the response of the OSWEC to the combined sea state. Figure \ref{fig:dias_sum} shows two separate outputs, angular rotation ($\boldsymbol{\theta}$) in the second row, and hydrodynamic torque ($\boldsymbol{\tau}_{h}$) in the third row. The responses to the combined sea state show that some OSWEC parameters are more affected by higher frequency excitation than others. For example, the effect of the wind wave component is clearly apparent in hydrodynamic torque, $\boldsymbol{\tau}_{h}$, but not visible in flap position, $\boldsymbol{\theta}$.

\subsubsection{WEC-Sim}
\label{sec:wecsim}
The second modeling tool we use to generate data is the open-source WEC modeling code WEC-Sim~\cite{wecsim}. WEC-Sim is a mid-fidelity program implemented in MATLAB that solves the six degree-of-freedom equation of motion in the time domain:
\begin{equation}
    \boldsymbol{M}\Ddot{\boldsymbol{y}}(t) = \boldsymbol{F}_{e}(t) + \boldsymbol{F}_{rad}(t) + \boldsymbol{F}_{b}(t) + \boldsymbol{F}_{\nu}(t) + \boldsymbol{F}_{md}(t) + \boldsymbol{F}_{PTO}(t),
    \label{eqn:WECSimEOM}
\end{equation}
where $\boldsymbol{M}$ is the mass matrix of the system, $\boldsymbol{F}_{e}$ is the excitation force vector, $\boldsymbol{F}_{rad}$ is the radiation force vector, $\boldsymbol{F}_{b}$ is the buoyancy force vector, $\boldsymbol{F}_{\nu}$ is damping force vector, $\boldsymbol{F}_{md}$ is the mean drift force vector, $\boldsymbol{F}_{PTO}$ is the control (PTO) force vector, and $\boldsymbol{y}$ is the six degree-of-freedom translation vector. WEC-Sim uses a boundary element method (BEM) solver based on linear potential flow theory to determine the hydrodynamic parameters, including $\mu$, $\nu$ and $\boldsymbol{F}_{e}$, then solves Equation \ref{eqn:WECSimEOM} in the time domain. 

We utilize built-in WEC-Sim capabilities to model weakly nonlinear OSWEC behavior in regular and irregular seas (Section \ref{sec:nonlinear} and \ref{sec:irregular}, respectively). WEC-Sim generates the weakly nonlinear behavior by calculating the instantaneous water surface elevation on the flap, and using that information to calculate the excitation and buoyancy forces, $\boldsymbol{F}_{e}$ and $\boldsymbol{F}_{b}$, respectively. WEC-Sim also allows for the use of irregular wave spectrum as an input to the system. We consider OSWEC dynamics in response to both a regular wave in Section \ref{sec:nonlinear}, and a real wave spectrum generated from field data in Section \ref{sec:irregular}. Finally, we note that, for equivalent wave inputs, the semi-analytical model and WEC-Sim are in close agreement (\ref{sec:diaswecsimcompare}).

\subsection{Preprocess data}

Regardless of the data source, the next step is to preprocess the data prior to running the DMD algorithm (Block 2 of Figure \ref{fig:workflow}). This begins with choosing which measurements to include in the DMD algorithm. Depending on the capabilities of the two data generation tools listed above, we do not use all of the states listed in Figure \ref{fig:oswec_system} for each case we consider. However, the number of states we include determines the rank of our data matrix (as described in Section \ref{sec:dmd}), which needs to be large enough compared with the underlying rank of the system dynamics. Because of this, in cases with more complex dynamics, it may be necessary to include more states or employ a DMD variant with time delays (as described in Section \ref{sec:timedelays}). For example, we could use four states, as indicated by the green boxes in Figure \ref{fig:workflow}, or we could only include the two kinematic states, $\boldsymbol{\theta}$ and $\dot{\boldsymbol{\theta}}$, if we thought those states were the only ones necessary to describe the dynamics of the system. The number of states and which states to include may not be possible to know a priori, but is critical to getting accurate results using DMD. Therefore, care must be taken to include enough dynamically-rich states to be able to describe the system under consideration.

Next, we normalize each state time series by its root mean square (RMS) value. For states with the same units (such as $\boldsymbol{F}_{x}$ and $\boldsymbol{F}_{z}$), we normalize each time series by the maximum RMS of all of the states with those units. This normalization step scales the relative importance of states of similar units, and eliminates the significant difference in magnitude between states with different units. For example, angular rotation and hydrodynamic torque can easily differ in magnitude by three orders or more, which can significantly affect the accuracy of the DMD algorithm without normalization. Normalizing data in general is a common step in many DMD studies to promote DMD accuracy and stability~\cite{proctor2015discovering,mohan2018data,serani2023use}.  

The next step is optional and we only use it in certain cases. The first option is to add artificial noise to the state measurements, which we use in Section \ref{sec:noise} to test robustness. There, we add white Gaussian noise with a defined signal-to-noise (SNR) ratio to approximate the noise present in physical systems. The second option is augment the available states using time delays~\cite{Brunton2017natcomm}, which we use for Section \ref{sec:nonlinear}. This process is discussed in more detail in Section \ref{sec:timedelays}. 

The final step of preprocessing is partitioning the state time series into training and testing data. The training data is the input to the DMD algorithm, and the testing data is used to assess how well DMD can predict future state behavior. 

\subsection{Dynamic mode decomposition}
\label{sec:dmd}
After preprocessing, we input the training data into the DMD algorithm. This section includes a brief derivation of the exact DMD algorithm~\cite{tu2013dynamic,kutz2016dynamic} with a detailed derivation included in Appendix \ref{sec:dmdappendix}. There are many variants~\cite{kutz2016dynamic,Brunton2022siamreview} to the DMD algorithm, some of which are also summarized here.  

The first step of the DMD algorithm is to arrange the training data into data matrices of the form: 
\begin{equation}
\boldsymbol{X} = 
\begin{bmatrix}
| & | & & | \\
\boldsymbol{x}_{1} & \boldsymbol{x}_{2} & ... & \boldsymbol{x}_{m} \\
| & | & & | \\
\end{bmatrix}, \hspace{0.3cm} \boldsymbol{X}^{'} = 
\begin{bmatrix}
| & | & & | \\
\boldsymbol{x}_{2} & \boldsymbol{x}_{3} & ... & \boldsymbol{x}_{m+1} \\
| & | & & | \\
\end{bmatrix} 
\hspace{0.2cm} \in \mathbb{R}^{nxm},
\end{equation}
where $\boldsymbol{x}_{k}$ is a column vector of all the states at time $t_{k} = k \Delta t$ and $\Delta t$ is the time step of the data. In these data matrices, each state vector time series corresponds to one \textit{row} of $\boldsymbol{X}$, while each \textit{column} is a snapshot of \textit{all} states at the associated time step. Note that the two data matrices $\boldsymbol{X}$ and $\boldsymbol{X}^{'}$ are shifted by a single time step. We can then define a best-fit linear operator, $\boldsymbol{A}$, that relates the time-evolution of our two data matrices:
\begin{equation}
    \boldsymbol{X}^{'} \approx \boldsymbol{AX},
    \label{eqn:regression}
\end{equation}
 where $\boldsymbol{A}$ relates the states of the OSWEC system from one time step to the next. In other words, $\boldsymbol{A}$ represents the system dynamics, and if we can approximate $\boldsymbol{A}$, we can approximate past and future system behavior. To solve for $\boldsymbol{A}$, we use the single value decomposition (SVD) of the data matrix $\boldsymbol{X}$ to calculate its pseudoinverse, $\boldsymbol{X}^{\dag}$ such that:
\begin{equation}
    \boldsymbol{A} = \boldsymbol{X}^{'} \boldsymbol{X}^{\dag} = \boldsymbol{X}^{'} \boldsymbol{V} \boldsymbol{\Sigma}^{-1} \boldsymbol{U}^{*},
\end{equation}
where $\boldsymbol{X} = \boldsymbol{U \Sigma V}^{*}$. In other words, the columns of $\boldsymbol{U}$ and $\boldsymbol{V}$ are the left and right singular vectors of $\boldsymbol{X}$, respectively, and $\boldsymbol{\Sigma}$ contains the associated singular values. Often, for high-dimensional data,  $\boldsymbol{A}$ is projected onto a low-dimensional subspace spanned by the first few columns of $\boldsymbol{U}$. This is described in detail in Appendix \ref{sec:dmdappendix}.

The eigenvectors of $\boldsymbol{A}$ are the DMD modes, $\boldsymbol{\phi}_{j}$, with corresponding discrete-time eigenvalues $\lambda_j$ that represent their growth/decay rates and frequencies. Using the DMD eigenvectors and eigenvalues, we can describe the system dynamics:
\begin{equation}
    \boldsymbol{x}(t) \approx \sum_{j=1} \boldsymbol{\phi_{j}} \text{exp}(\gamma_{j} t) b_{j} = \boldsymbol{\Phi} \text{exp}(\boldsymbol{\Gamma} t) \boldsymbol{b},
    \label{eqn:dynamics}
\end{equation}
 where the columns of $\boldsymbol{\Phi}$ contain the DMD modes $\boldsymbol{\phi}_{j}$, $\gamma_{j} = {\text{ln}(\lambda_{j})}/{\Delta t}$ are the complex continuous-time DMD eigenvalues, and $\boldsymbol{b} = \boldsymbol{\Phi}^{\dag} \boldsymbol{x}_{1}$ are the DMD mode amplitudes with $\boldsymbol{x}_{1}$ being the states at time $t_{1}$. Note that $\boldsymbol{x}(t)$ contains the time series all state variables. With Equation \ref{eqn:dynamics}, we can now describe and predict system state dynamics using only snapshot data of our system. 

\subsubsection{Total-least-squares DMD}
\label{sec:TLSDMD}
One of DMD’s significant limitations is its sensitivity to noise~\cite{duke2012error,bagheri2014effects}. In particular, noise biases eigenvalues in a way that can cause an artificial exponential decay in DMD dynamics~\cite{bagheri2014effects,dawson2016characterizing}. These effects cannot be counteracted by increasing the duration or sampling frequency of data considered~\cite{hemati2017biasing}. To remedy this, Hemati et al. developed a  modification to the DMD algorithm called total-least-squares DMD (TLS DMD) that uses a total-least-squares approach to solve the regression problem laid out in Equation \ref{eqn:regression}, rather than the traditional least-squares method~\cite{hemati2017biasing}. By doing this, this algorithm assumes noise is present in both data matrices, rather than just $\boldsymbol{X}$, as in the original algorithm. The mathematical derivation of the TLS DMD algorithm is described in Appendix \ref{sec:TLSDMDappendix} and we use this variation in Section \ref{sec:noise}.

\subsubsection{Time delays}
\label{sec:timedelays}
As previously mentioned, another variation of DMD involves the use of ``time delays''. Tu et al. \cite{tu2013dynamic} were the first to point out that using time-shifted matrices (i.e., time delay coordinates or Hankel matrices) can improve the performance of DMD, particularly on periodic data, by providing additional phase information of the dynamics. The use of time delays was later connected to the Koopman operator as a universal linearizing basis~\cite{Brunton2017natcomm}. Details on using time delays are provided in Appendix \ref{sec:TDappendix} and we use this variation in Section \ref{sec:nonlinear}. 

\subsubsection{Optimized DMD}
\label{sec:optDMD}
The last variation of DMD we use in this study is optimized DMD (optDMD), as described in Askham and Kutz~\cite{askham2018variable}. Instead of solving a regression problem for linear operator $\boldsymbol{A}$ using two time-shifted data matrices as shown in Equation \ref{eqn:regression}, optDMD aims to fit the data matrix $\boldsymbol{X}$ directly to a series of exponential functions, $e^{\gamma_{j}t}$, where $\gamma$ are the continuous-time DMD eigenvalues. The algorithm then uses variable projection techniques to efficiently solve the resulting nonlinear least squares problem to determine the weight of each exponential, which can be used to calculate the DMD eigenvectors. One main difference between optDMD and exact DMD is optDMD simultaneously optimizes DMD modes and amplitudes using the full data set, rather than finding the DMD modes, and then solving for DMD mode amplitudes using only the first data snapshot like exact DMD. As a result, optDMD provides an optimized DMD model that can handle non-ideal data and has less bias for noisy inputs. This algorithm also allows for the addition of constraints on DMD eigenvalues that restrict the real part of the continuous-time eigenvalues to be approximately zero, which limits the artificial temporal decay and promotes stability in the DMD reconstruction. We use optDMD in Section \ref{sec:irregular} to model OSWEC behavior in response to irregular waves.  We provide a brief derivation of optDMD in Appendix \ref{sec:optDMDappendix}, and refer the reader to~\cite{askham2018variable} for a full derivation as well as open-source code.

\subsection{Evaluate performance}
Finally, the last step in the workflow (Block 4 in Figure \ref{fig:workflow}) is to evaluate the performance of the DMD model. We evaluate the accuracy with which the DMD model describes the training data (hindcasting) and predicts the testing data (forecasting). For this evaluation, we use a normalized error parameter $\epsilon$ that can be calculated for each state variable, $\boldsymbol{\psi}$, corresponding to one row in $\boldsymbol{X}$. We define $\epsilon$ as the ratio between the $L_2$ norm of the difference between the DMD output, $\boldsymbol{\psi}_{DMD}$, and the true state variable $\boldsymbol{\psi}$, normalized by the $L_2$ norm of $\boldsymbol{\psi}$:
\begin{equation}
\label{eqn:epsilon}
    \epsilon = \frac{ \left| \left| \boldsymbol{\psi} - \boldsymbol{\psi}_{DMD} \right| \right|_{2}}{\left| \left| \boldsymbol{\psi} \right| \right|_{2}}.
\end{equation}
This error parameter can be applied to the training or testing region, where we can replace $\boldsymbol{\psi}$ for $\boldsymbol{\psi}_{train}$ or $\boldsymbol{\psi}_{test}$, respectively, and do a similar replacement for $\boldsymbol{\psi}_{DMD}$.

\subsection{Test cases}

We consider three cases to evaluate how well DMD can model realistic OSWEC behavior. The relevant parameters used in each case are outlined in Table \ref{tab:dmd_cases}. 

The first case evaluates how well DMD can model OSWEC dynamics when noise is present in the training data. Because any real WEC system will have some level of sensor noise, it is critical that our modeling method can account for that and still model the system behavior well. For this case, we use the semi-analytical model to generate the training data and a polychromatic wave input composed of two wave components. We artificially add Gaussian white noise to the data after normalization (Block 2 in Figure \ref{fig:workflow}) at different signal-to-noise ratios (SNRs) and evaluate the accuracy of exact DMD and total-least-squares DMD, given this input. We use 10 seconds of training data, and consider a testing region lasting 30 seconds. Finally, we include states $\boldsymbol{\theta}$, $\dot{\boldsymbol{\theta}}$, $\boldsymbol{\tau}_{h}$, and three pressure sensors. All this information is summarized in the first column of Table \ref{tab:dmd_cases}.

The second case evaluates how well DMD, a linear algorithm, can model weakly nonlinear dynamics. We use WEC-Sim's capability to model the flap as a weakly nonlinear body and evaluate the accuracy of exact DMD with and without time-delays. Similarly to Case 1, we use 10 seconds of training data and consider a 30 second testing region. We include states $\boldsymbol{\theta}$, $\boldsymbol{\dot{\theta}}$, $\boldsymbol{\tau}_{h}$, $\boldsymbol{F}_{x}$, and three pressure sensors. Refer to the second column in Table \ref{tab:dmd_cases} for a summary of parameters used in Case 2. 

For the third case, we evaluate how well DMD can model OSWEC behavior in response to an irregular wave forcing. We input a wave spectrum calculated from field data into WEC-Sim and model the weakly nonlinear OSWEC output. Unlike the other cases, we calculate the power spectra of our state variables over time and use DMD to model the temporal behavior of the power spectra, rather than the time evolution of the state variables, which requires a slight change in preprocessing. We expect the resulting dynamics to be more complex and high-dimensional than the previous two cases, which can result in unstable and inaccurate results from exact DMD, especially with a limited number of system states. Because of this, we use optDMD in addition to exact DMD and compare their performance. In addition, because of the increased complexity we use a significantly longer training time of 10 minutes (600 seconds), and consider a testing region of 5 minutes (300 seconds) to ensure there is enough training time to capture the structure and periodicity in the states' spectra. Finally, we include only four states: $\boldsymbol{\theta}$, $\dot{\boldsymbol{\theta}}$, $\boldsymbol{\tau}_{h}$, and $\boldsymbol{F}_{x}$. These parameters are summarized in the third column of Table ~\ref{tab:dmd_cases}.

In summary, we chose these three cases as examples not only because they specifically address common barriers in WEC modeling and control, but also because these are cases where we expect the unmodified DMD algorithm to perform poorly. We then demonstrate that the different extensions of DMD can overcome each of these complications (e.g. noise, nonlinearity, latent variables). By highlighting these cases, we show that DMD can be an effective tool at modeling realistic WEC dynamics.

\begin{table}[]
\label{tab:dmd_cases}
\centering
\caption{Summary of data parameters and algorithms used for the three cases. DMD refers to the exact DMD algorithm and serves to contrast with variants including total-least-squares DMD (TLS DMD), DMD with time delay (DMD+TD), and optimized DMD (optDMD).}
\begin{tabular}{|c|c|c|c|}
\hline
                     & \textbf{Case 1}       & \textbf{Case 2} & \textbf{Case 3} \\
\hline
\textbf{Barrier}     & Measurement noise     & Nonlinearity   & Irregular waves \\ 
\hline
\textbf{Data source} & Semi-analytical model & WEC-Sim        & WEC-Sim \\
\hline
\textbf{Wave type}   & Polychromatic         & Monochromatic  & Irregular \\
\hline
\textbf{\begin{tabular}[c]{@{}c@{}}Wave\\  parameters\end{tabular}} & \begin{tabular}[c]{@{}c@{}}$H_{\text{swell}}$ = 0.75 m, $H_{\text{wind}}$ = 0.3 m\\ $T_{\text{swell}}$ = 8 s, $T_{\text{wind}}$ = 2.55 s\end{tabular} & \begin{tabular}[c]{@{}c@{}}H = 2 m\\ T = 8 s\end{tabular}  & \begin{tabular}[c]{@{}c@{}}$H_{s}$ = 0.9 m\\ $T_{p}$ = 5.8 s\end{tabular}        \\ 
\hline
\textbf{Algorithms} & \begin{tabular}[c]{@{}c@{}}DMD\\ TLS DMD\end{tabular}  & \begin{tabular}[c]{@{}c@{}}DMD\\ DMD+TD\end{tabular} & \begin{tabular}[c]{@{}c@{}}DMD\\ optDMD\end{tabular}                       \\ 
\hline
\textbf{Training time}  & 10 s & 10 s  & 600 s   \\ 
\hline
\textbf{Testing time}   & 30 s & 30 s  & 300 s \\ 
\hline
\textbf{States} & \begin{tabular}[c]{@{}c@{}}$\boldsymbol{\theta}$\\ $\dot{\boldsymbol{\theta}}$\\ $\boldsymbol{\tau}_{h}$\\ $\boldsymbol{P}_{1} - \boldsymbol{P}_{3}$\end{tabular}  & \begin{tabular}[c]{@{}c@{}}$\boldsymbol{\theta}$\\ $\dot{\boldsymbol{\theta}}$\\ $\boldsymbol{\tau}_{h}$\\ $\boldsymbol{F}_{x}$\\ $\boldsymbol{P}_{1} - \boldsymbol{P}_{3}$\end{tabular} & \begin{tabular}[c]{@{}c@{}}$\boldsymbol{\theta}$\\ $\dot{\boldsymbol{\theta}}$\\ $\boldsymbol{\tau}_{h}$\\ $\boldsymbol{F}_{x}$\end{tabular} \\
\hline
\end{tabular}
\end{table}

\section{Results and Discussion}
\label{sec:results}

\subsection{Case 1: Noisy measurements}
\label{sec:noise}

\begin{figure}[t!]
    \centering
    \includegraphics[width=1\linewidth]{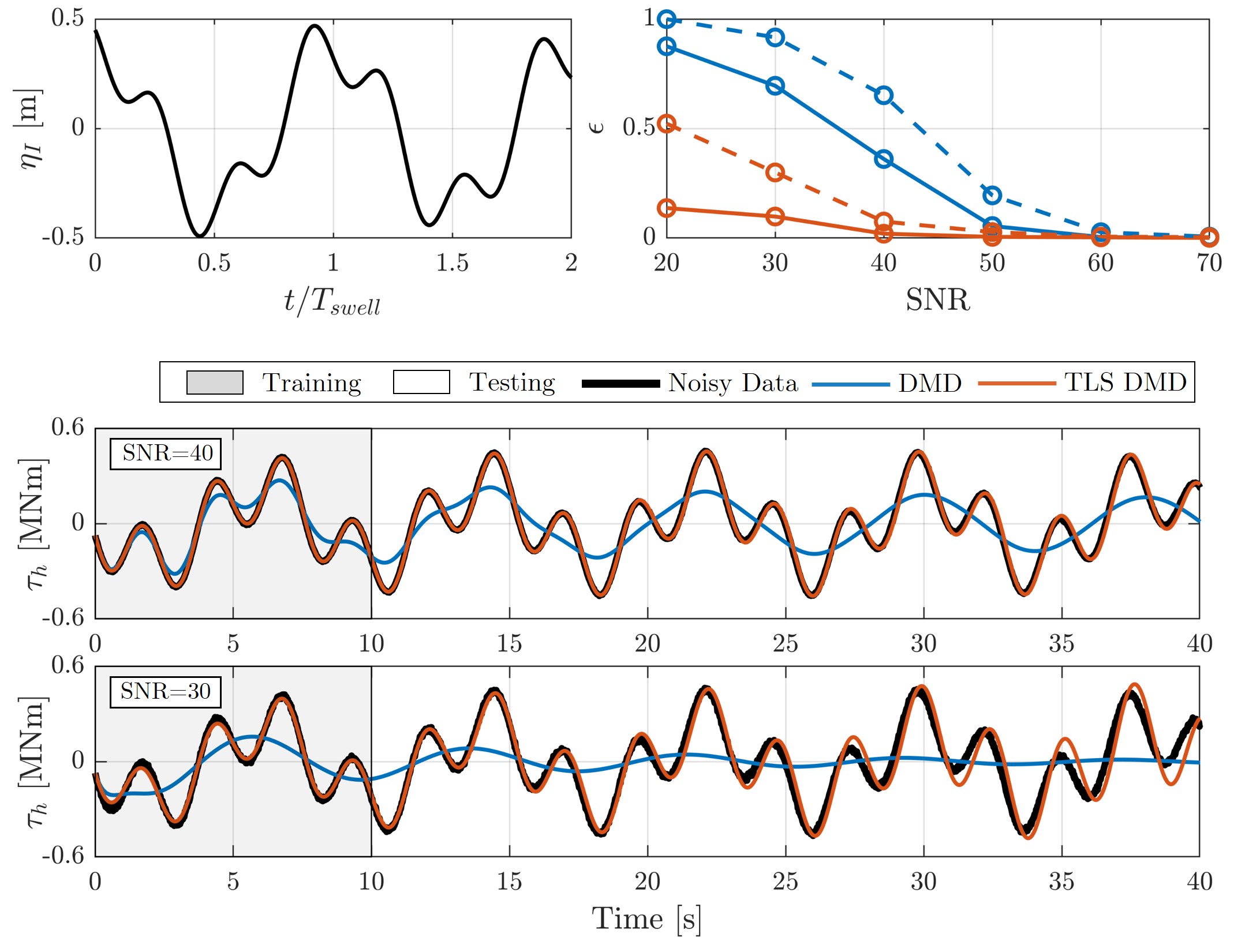}
    \caption{(top left) Undisturbed incident wave field at the flap over time normalized by the wave period of the swell wave component, $T_{\mathrm{swell}}$. (top right) Normalized error of DMD output, $\epsilon$, for state variable $\boldsymbol{\tau}_{h}$ as a function of SNR. Lower values of SNR correspond to a noisier signal. Dashed lines represent errors in the training region (hindcasting) and solid lines represent error in the testing region (forecasting). As SNR decreases, TLS DMD (orange) consistently outperforms exact DMD (blue) in both the testing and training region. (bottom panels) Hydrodynamic torque from WEC-Sim (black), exact DMD (blue), and TLS DMD (orange) for SNR value of 40 (top) and SNR of 30 (bottom). The time series show the artificial decay in the DMD model, but much better accuracy for TLS DMD, consistent with the normalized error values.}
    \label{fig:noisy1}
\end{figure}

Figure \ref{fig:noisy1} summarizes the DMD and TLS DMD performance on OSWEC data augmented with white Gaussian noise. The top left panel shows the undisturbed incident wave field located at the flap ($x = 0$) as a function of time normalized by the period of the swell wave component ($T_{\mathrm{swell}} = 8$ seconds). The swell and wind wave components in the incident wave are clearly visible. The top right panel shows normalized error, $\epsilon$, for hydrodynamic torque, $\boldsymbol{\tau}_{h}$, as a function of SNR, with 70 being the least amount of noise added, and 30 being the most. Error from the exact DMD algorithm is shown in blue, while that of the TLS DMD algorithm is shown in orange. Although error rises monotonically with increasing noise for both training (solid line) and testing (dashed line) data, TLS DMD consistently outperforms exact DMD. We observe these trends for all states (not shown). In experimental setups, it is possible that noise levels could exceed those tested in this study. In those cases, a different DMD variant, such as optDMD~\cite{askham2018variable}, may better address the issue of noise bias for lower SNR values at a slightly higher computational cost.

The two time series panels at the bottom of Figure \ref{fig:noisy1} show hydrodynamic torque, $\boldsymbol{\tau}_{h}$ with added white noise with an SNR of 40 (top) and 30 (bottom). The black lines represent the semi-analytical model output augmented with white Gaussian noise, the blue line represents the dynamics described by the exact DMD algorithm, and the orange line represents the dynamics described by TLS DMD. The training duration (10 seconds) is represented by the gray highlighted region. For both time series, exact DMD artificially decays through the test region. This is common behavior of DMD models using noisy data~\cite{bagheri2014effects}. However, the TLS DMD is able to capture higher frequency oscillations from wind waves without significant decay. The TLS prediction accuracy does decrease over time, which is most notable in the SNR = 30 case, but this becomes less of an issue when considering the application of WEC control, where this model would be used to forecast the system behavior over a finite-time horizon (e.g., a few wave periods into the future).

\begin{figure}[t!]
    \centering
    \includegraphics[width=1\linewidth]{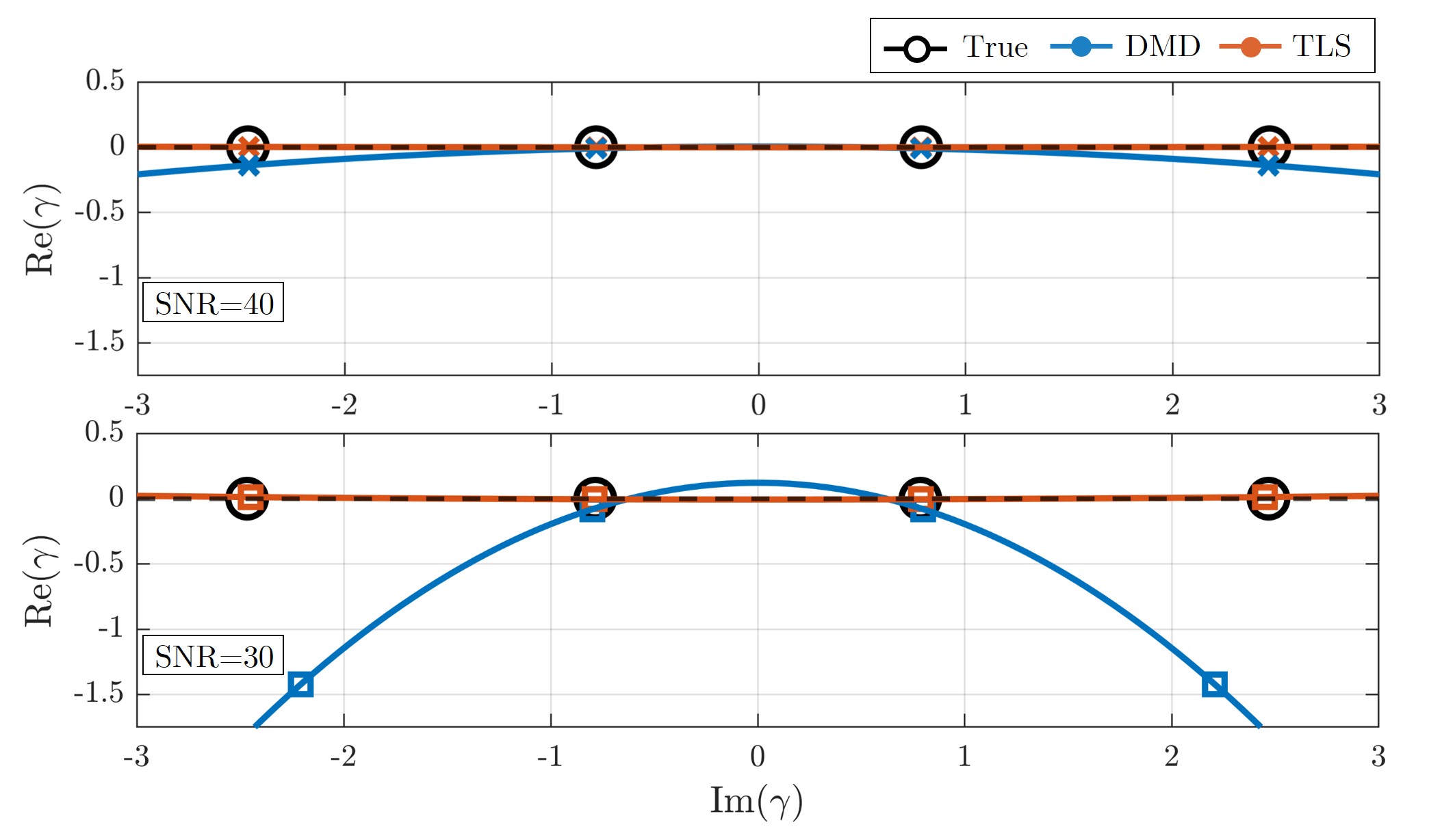}
    \caption{Continuous-time DMD eigenvalues, $\gamma$, on the complex plane with best-fit parabolic curves. High curvature of the eigenvalues represent a decrease in the real part of the eigenvalues, which in turn result in higher exponential decay in the time series forecasts, as characterized in~\cite{bagheri2014effects}. Black open circles are the true DMD eigenvalues of the system, the blue symbols represent exact DMD eigenvalues, and the orange symbols represent TLS DMD. The x markers represent the SNR=40 case (top panel), while the square markers represent the SNR=30 case (bottom panel). As SNR increases, the curvature of the best-fit lines for exact DMD increases significantly, which corresponds to the exponential decay seen in the two time series in Figure \ref{fig:noisy1}. In contrast, the best-fit curve for the TLS DMD eigenvalues (orange) show almost no curvature, which corresponds to significantly less error in the time series fits.}
    \label{fig:noisy2}
\end{figure}

For further insight into the differential performance of DMD and TLS DMD, we consider the continuous-time DMD eigenvalues, $\gamma_{j} \in \mathbb{C}$, in the complex plane. Bagheri~\cite{bagheri2014effects} showed that adding noise to a system decreases the real component of the DMD eigenvalues and this effect is amplified for higher harmonics of the eigenvalue frequency. This results in a parabolic behavior of DMD eigenvalues in the complex plane, with higher curvature corresponding to higher levels of noise. In addition, a decrease in the real part of the DMD eigenvalue results in an exponential decay in the DMD approximation of the time dynamics, which can be seen from Equation \ref{eqn:dynamics}. Figure \ref{fig:noisy2} visualizes this pattern for the SNR=40 (top) and SNR=30 (bottom) cases with the DMD and TLS DMD eigenvalues plotted in the complex plane. The black open circles represent the DMD eigenvalues of the noise-free system, which all have a real part of zero, representing a stable system. The exact DMD eigenvalues are shown in blue, and the TLS DMD eigenvalues are shown in orange, with corresponding best-fit parabolic curves. As expected, we see a significant increase in curvature with an increase in the noise amplitude (corresponding to a decrease in SNR), but we also see the dramatic reduction in curvature of the TLS DMD eigenvalues in both cases. This reduction in curvature means TLS DMD has significantly less spurious decay in time and does a much better job at modeling the noisier system dynamics than the exact DMD algorithm. Although we used six states in our data matrix, we chose to use to include only four SVD modes in the DMD reconstruction, which results in four eigenvalues in Figure \ref{fig:noisy2}. We did this because we knew for an idealized system with two frequency components, we need two pairs of complex conjugates for eigenvalues to describe the dynamics \cite{tu2013dynamic}. It is common to run DMD using less SVD modes than you have available to make the algorithm more efficient \cite{kutz2016dynamic}, and details of that are described in Appendix \ref{sec:dmdappendix}. In practice, it is likely the rank of the system cannot be known a priori, so the number of SVD modes to include is a free variable that may need adjusting to optimize accuracy and run time for the algorithm.

These results show that while exact DMD may not be suitable for systems with even low levels of noise, TLS DMD is an effective tool for modeling noisy data and can substantially improve the accuracy of the models in both the training and testing regions. This is promising because it shows that we have an efficient and accurate method to model realistic OSWEC systems with only a slight modification to the original DMD algorithm. 

\subsection{Case 2: Nonlinear behavior}
\label{sec:nonlinear}

Next we explore the potential for DMD, a linear algorithm, to model weakly nonlinear dynamics. Although DMD uses linear least-squares regression, its relation to the Koopman operator means it can be useful to model nonlinear systems~\cite{rowley2009spectral,mezic2013analysis,Brunton2022siamreview,colbrook2022mpedmd,colbrook2023residual}. Here, we summarize the ability of exact DMD and DMD with a time delay to model the dynamics of a weakly nonlinear OSWEC in response to a regular wave.  

\begin{figure}[t!]
    \centering
    \includegraphics[width=1\linewidth]{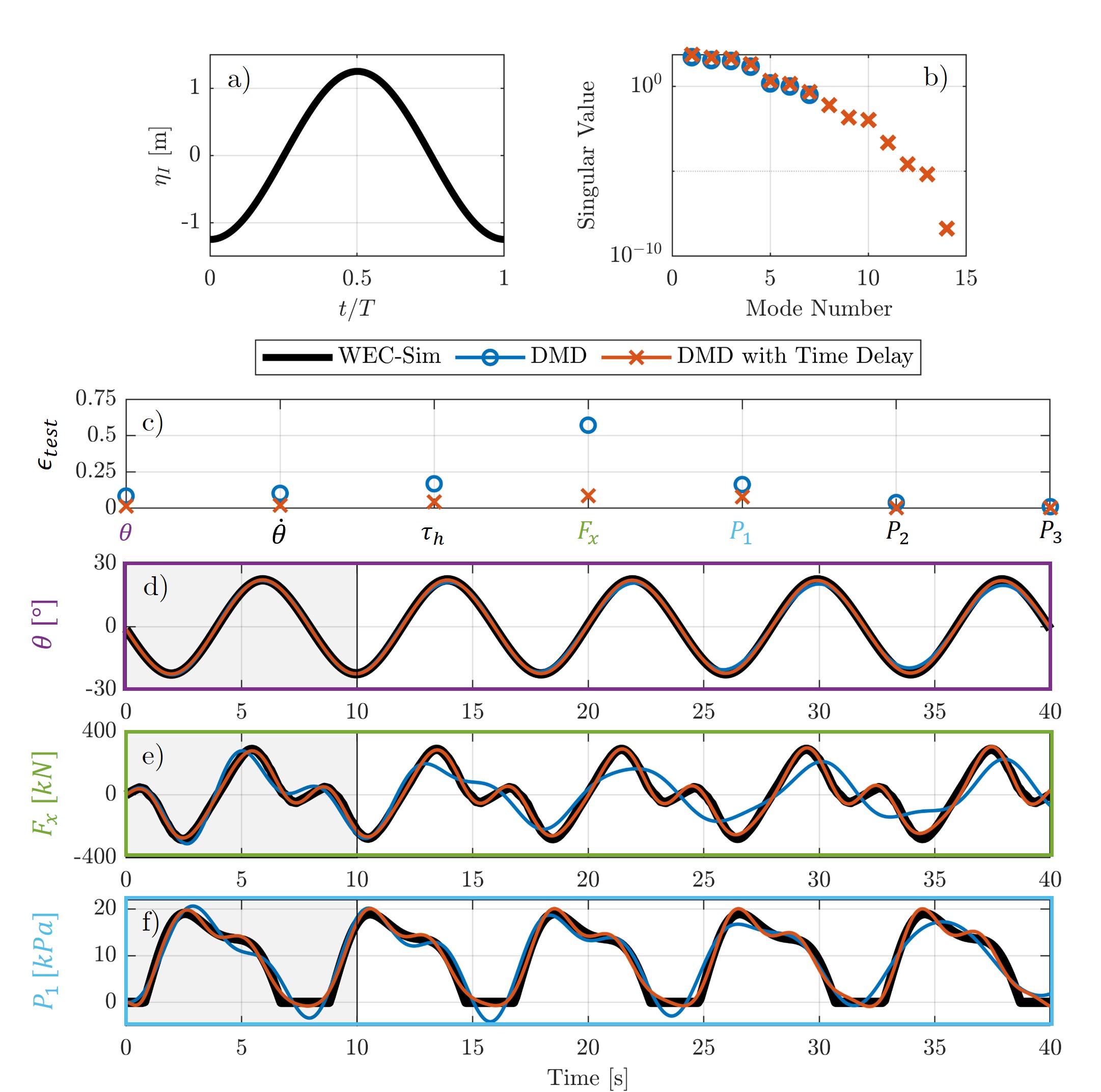}
    \caption{(a) Regular wave input over one oscillation period. (b) Singular values of data matrix $\boldsymbol{X}$ from exact DMD (blue circles) and DMD with one time delay (orange crosses). The singular values for the time delay case show that the true system is a higher rank (contains important information on dynamics in higher modes) than can be captured with the exact DMD case. (c) Normalized testing error, $\epsilon_{test}$, as a function of system states, showing that DMD with time delay (orange crosses) outperforms exact DMD (blue circles) for all states with nonzero error. Time series of (d) angular rotation, $\boldsymbol{\theta}$, (e) surge force, $\boldsymbol{F}_{x}$, and (f) the top pressure sensor, $\boldsymbol{P}_{1}$. WEC-Sim output is in black, the exact DMD fit is in blue, the TLS DMD output is in orange, and the training region is highlighted in gray. By adding a time delay, the model captures higher order oscillations with better accuracy for $\boldsymbol{F}_{x}$ and $\boldsymbol{P}_{1}$, and performs better over time in the testing region for all three states shown.}
    \label{fig:nonlinearity}
\end{figure}

Figure \ref{fig:nonlinearity}(a) shows the time series of wave elevation over one wave period. The incident wave is a regular wave with a height of 2.5 meters and period of 8 seconds. Figure \ref{fig:nonlinearity}(b) describes the singular values of data matrix $\boldsymbol{X}$ from the exact DMD algorithm (blue circles) and DMD with one time delay (orange crosses). The number of singular values corresponds to the number of rows in the data matrix $\boldsymbol{X}$, and because adding a time delay doubles the rows in the data matrix, there are twice as many singular values for the time delay case. The decay of the singular values can give insight into the data. For example, if the decay is steep, there may be an obvious cutoff for modes needed for an accurate reconstruction of the data. On the other hand, if the decay is slow, then we may  need to include all modes in the reconstruction because truncation can leave out important dynamics. Figure \ref{fig:nonlinearity}(b) shows the singular values for the time delay case are less steep than for the exact DMD case, implying that more SVD modes are required for an accurate reconstruction. Therefore, adding time delays enables us to recover more information about dynamics that are otherwise hidden and latent for exact DMD. 

Figure \ref{fig:nonlinearity}(c) shows the normalized testing error parameter, $\epsilon_{test}$, as a function of system states, again with blue circles representing the error in the testing region for exact DMD, and the orange crosses representing testing error for DMD with a single time delay. Some states have near zero error for both cases (such as $\boldsymbol{P}_{2}$ and $\boldsymbol{P}_{3}$), but in every case with nonzero error, DMD with time delay outperforms exact DMD. 

Figure \ref{fig:nonlinearity}(d)-(f) shows the time series of selected states with the true dynamics shown in black, the output from exact DMD shown in blue, and the output from DMD with one time delay shown in orange. Figure \ref{fig:nonlinearity}(d) shows the time series for angular rotation, $\boldsymbol{\theta}$, with the training region shown in gray. Both DMD and DMD with a time delay model the time series well, but even in this case where the dynamics appear quite linear and monotonic, there is a slight decay in the exact DMD result without a time delay, meaning we are losing accuracy as we move further from the training region. However, that decay does not appear when adding a time delay, as shown in Figure \ref{fig:nonlinearity}(d) and indicated with a near-zero testing error in Figure \ref{fig:nonlinearity}(c). Figure \ref{fig:nonlinearity}(e) shows the time series for surge force acting on the flap, $\boldsymbol{F}_{x}$, which contains more interesting dynamics. This parameter is most improved when adding a time delay, as shown in (c). This is likely because without a time delay, there are not enough SVD modes available to fully capture the higher order oscillations in the time series. In addition, the accuracy of the DMD model without a time delay degrades over time. However, adding a time delay results in near perfect agreement for the same amount of training time, as shown by the testing error being below 0.1. Figure \ref{fig:nonlinearity}(f) shows a similar trend for the pressure measurement near the top of the flap. The pressure measurements have no negative values, leading to a piece-wise behavior that can be difficult to model. Both DMD and DMD with a time delay do a good job of modeling the general dynamics of the pressure data, but DMD with a time delay is better able to describe the data, especially in later times in the testing period. 

These results demonstrate that adding a time delay can significantly increase the performance of DMD for weakly nonlinear systems. Even in cases where DMD provides a reasonably good description of the training data, adding a time delay can increase accuracy of higher order nonlinear patterns well into the testing region that can be vital for predicting future performance far beyond the training region.

\subsection{Case 3: Irregular wave input}
\label{sec:irregular}

For this final case, we investigate how well DMD can model OSWEC behavior in response to a real wave spectrum taken from field data. Irregular waves raise a range of potential complications, as the process can be nonlinear, nonstationary, and stochastic. In other words, frequency content of the incident wave field changes over time, and there may be phenomena such as frequency wandering and nonlinear wave-to-wave interactions. This means that with limited sensor measurements and a broad frequency range of incident waves, DMD is not able to consistently model time-domain dynamics of the state variables with a reasonable number of sensors and training time, even using time delays. If an irregular wave spectrum was purely periodic (which is not the case in practice), it would be possible for DMD to model the time series similar to that of Cases 1 and 2, but with two major limitations. First, this would require significantly more modes, which would, in turn, require more sensor measurements due to the diminishing returns from multiple time delays. Second, the training time would need to encompass most of the repetition period, which raises concerns about computational cost and the general suitability of the method for control. 

Because of this, rather than using DMD to model and predict magnitudes of sensor measurements, as in Cases 1 and 2, we instead use DMD to model and predict the time behavior of the power spectral distributions (PSDs) of the state variables. The power spectrum of state $\boldsymbol{\psi}$ over a specified time window $T_{w}$ is described as:
\begin{equation}
    \boldsymbol{S}_{\boldsymbol{\psi}} = \frac{2\left| \text{FFT}({\boldsymbol{\psi}}\sp{\prime}) \right|^{2}}{m \hspace{0.05cm} f_{s}},
\end{equation}
 where $\boldsymbol{\psi}\sp{\prime}$ is the windowed and filtered time signal of state $\boldsymbol{\psi}$, $m$ is the number of time steps in the windowed signal, and $f_{s}$ is the sampling frequency. We apply a sliding window Hann filter to the state variable time series and look at the PSD of the filtered data, such that $\boldsymbol{S}_{\boldsymbol{\psi}}$ is a function of both frequency and time and represents how the spectral properties of the state vary in time. Each column in the spectrograms represents the PSD of 60 seconds of data.

\begin{figure}[t!]
    \centering
    \includegraphics[width=1\linewidth]{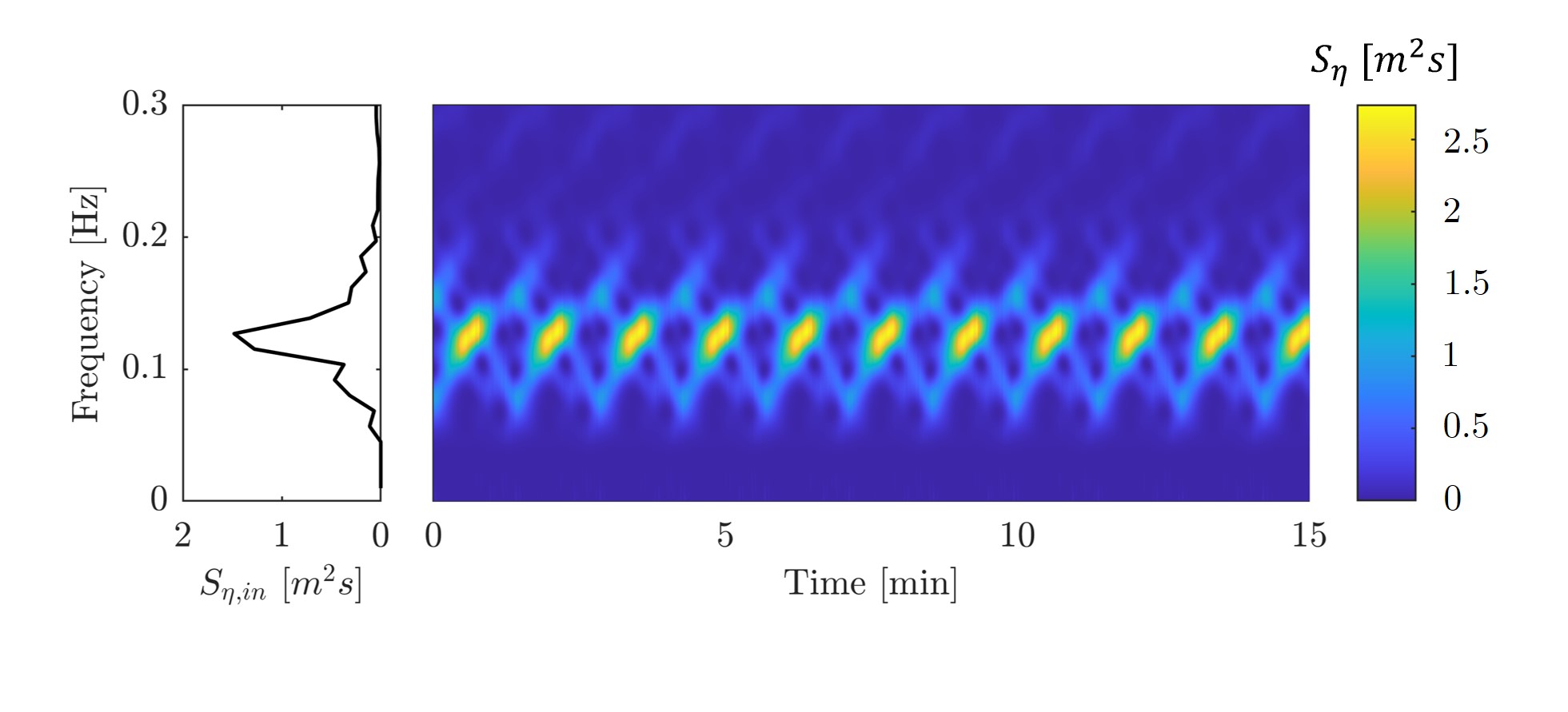}
    \caption{Periodogram (left) and spectrogram (right) of incident wave field used in Case 3. The periodogram is a direct measurement of average wave statistics from field data~\cite{thomson2012wave}, and was used as a wave input into WEC-Sim, $\boldsymbol{S}_{\boldsymbol{\psi},in}$. The spectrogram is of the resulting wave field time series generated by WEC-Sim. To generate the spectrogram, we use a sliding window short-time Fourier transform on filtered and zero-padded time series data using a Hann filter. We use a window width of 60 seconds and a window time step of 1 second to generate all spectrograms. We note that parameters such as filter type, window width, and window time step are tunable and may affect the accuracy of the results and DMD fits.}
    \label{fig:irreular_wave_spectrum}
\end{figure}

Consider the periodogram of the average wave power spectrum from field data taken off the coast of North Carolina using SWIFT drifters~\cite{thomson2012wave} (Figure \ref{fig:irreular_wave_spectrum}). The frequency distribution is relatively narrow, with a peak period of 5.8 s and a significant wave height of 0.9 m. WEC-Sim uses this spectrum to generate an irregular wave forcing in the time domain using a random phase seed for a set of frequencies, which we use to simulate a weakly nonlinear OSWEC response. The spectrogram of the resulting incident wave time series (Figure \ref{fig:irreular_wave_spectrum}) contains multiple frequencies with periodic structure. Although this exact repetition is unlikely in real field data, some seas can result in wave power spectra that are periodic and structured in time, especially in swell-dominated conditions~\cite{soares2005spectrogram}. Given the structure in the time behavior of the \textit{frequency} content of the incident wave, we expect similar structure to be present in system states that could be captured by a DMD model of the system state \textit{spectrograms}. In other words, we aim to recreate the spectrograms of state variables to model and predict their spectro-temporal behavior using DMD. Similar to Case 1 and 2, we partition the spectrogram for each state variable into a training and testing periods, the lengths of which are reported in Table \ref{tab:dmd_cases}. 

Because we are now using DMD to model and predict an evolving short-time Fourier transform (i.e., a spectrogram), the data matrices $\boldsymbol{X}$ and $\boldsymbol{X}^{'}$ are arranged differently. We stack the frequency data of each state in a single column at the corresponding time step. For example, if the state variables are $\boldsymbol{\theta}$ and $\dot{\boldsymbol{\theta}}$, the data matrix $\boldsymbol{X}$ is arranged as: 
\begin{equation}
\boldsymbol{X} = 
\begin{bmatrix}
| & | & & | \\
\boldsymbol{S}_{\boldsymbol{\theta},1} & \boldsymbol{S}_{\boldsymbol{\theta},2} & ... & \boldsymbol{S}_{\boldsymbol{\theta},m} \\
| & | & & | \\
\\
| & | & & | \\
\boldsymbol{S}_{\boldsymbol{\dot{\theta}},1} & \boldsymbol{S}_{\boldsymbol{\dot{\theta}},2} & ... & \boldsymbol{S}_{\boldsymbol{\dot{\theta}},m} \\
| & | & & | \\
\end{bmatrix}
,
\end{equation}
 where $\boldsymbol{S}_{\boldsymbol{\theta},1}$ would correspond to the first column of the spectrogram of state variable $\boldsymbol{\theta}$, $\boldsymbol{S}_{\boldsymbol{\theta},2}$ the second column, and so on. The rest of the DMD process is unchanged. In addition to exact DMD, we use optDMD (Section \ref{sec:optDMD}) to model the state variable spectrograms, and use a reconstruction rank of 30 modes for both algorithms. We compare their performance with a normalized error metric, $\Bar{\boldsymbol{\epsilon}}$, defined as:
\begin{equation}
    \Bar{\boldsymbol{\epsilon}} = \frac{\boldsymbol{S}_{\boldsymbol{\psi}}-\boldsymbol{S}_{\boldsymbol{\psi},DMD}}{\text{max}(\boldsymbol{S}_{\boldsymbol{\psi}})},
\end{equation}
 where $\text{max}(\boldsymbol{S}_{\boldsymbol{\psi}})$ is the maximum value of the true spectrogram of that state. Unlike the error parameter used in the previous two cases, $\Bar{\boldsymbol{\epsilon}}$ is a matrix of values the same size as the state spectrograms, not a singular value.

\begin{figure}[t!]
    \centering
    \includegraphics[width=1\linewidth]{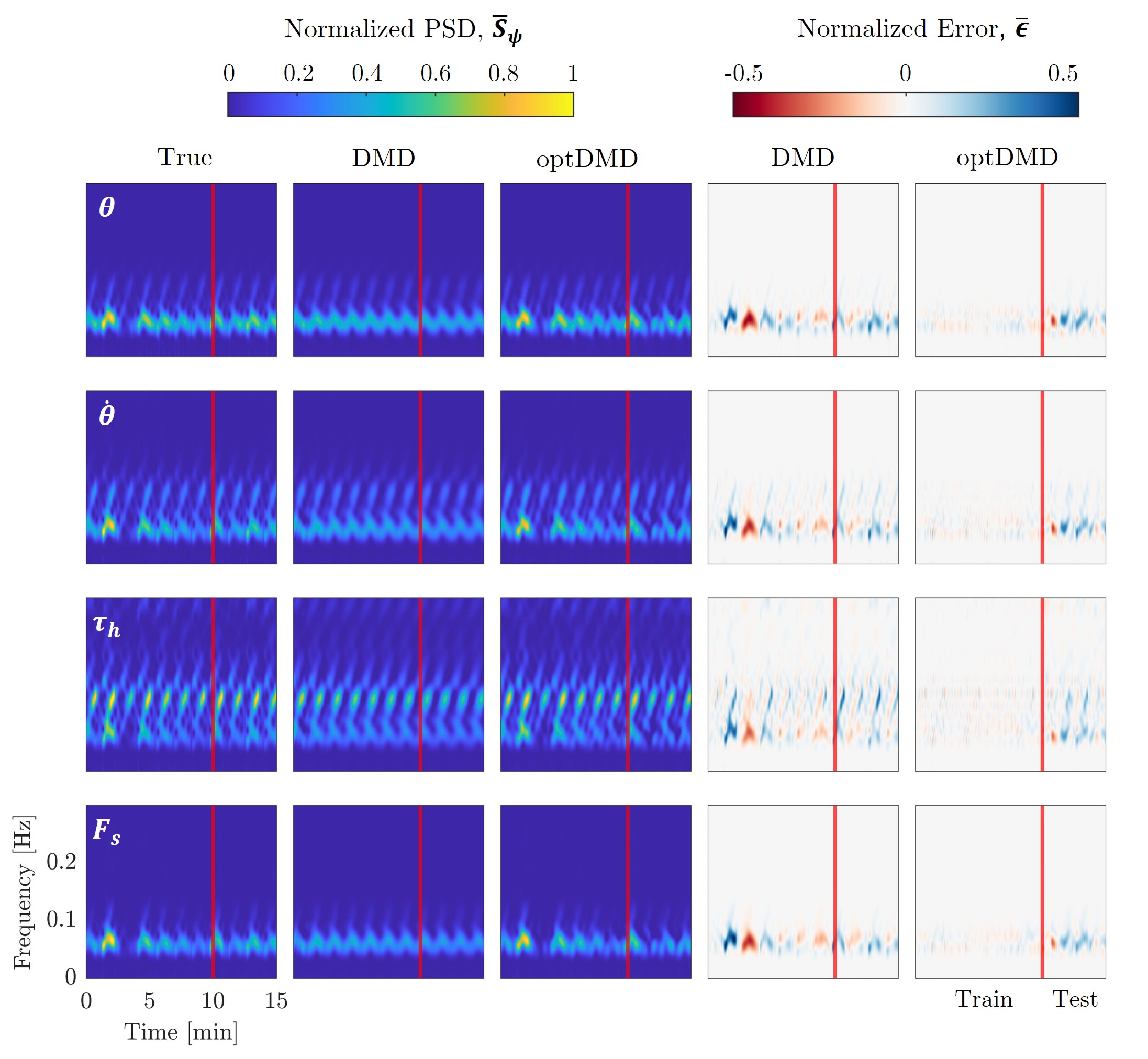}
    \caption{Spectrograms of WEC-Sim output (``True'') (first column), exact DMD (second column), and optDMD (third column) for state variables $\boldsymbol{\theta}$ (top row), $\dot{\boldsymbol{\theta}}$ (second row), $\boldsymbol{\tau}_{h}$ (third row), and $\boldsymbol{F}_{x}$ (bottom row). All spectrograms are normalized by the maximum value of the true spectrogram for the corresponding state variable. The last two columns represent the normalized error for the DMD model (fourth column) and optDMD fit (fifth column). optDMD outperforms DMD for all state variables, particularly in the training region, which ends at the red vertical line.}
    \label{fig:spectrogram_grid_r_14}
\end{figure}

Figure \ref{fig:spectrogram_grid_r_14} shows the DMD and optDMD models of the spectrograms for state variables: $\boldsymbol{\theta}$, $\dot{\boldsymbol{\theta}}$, $\boldsymbol{\tau}_{h}$, and $\boldsymbol{F}_{x}$. Each spectrogram is normalized is maximum value. There is evident temporal structure and periodicity in each of the ``true'' spectrograms (WEC-Sim data), which are features DMD should be able to capture. We note that, despite the strong repetition in the incident wave spectrogram (Figure \ref{fig:irreular_wave_spectrum}), the spectrograms of state variables contain more time-varying structure. Both DMD and optDMD can approximate the general spectro-temporal behavior of each state variable, but optDMD is more accurate than exact DMD in both the training and testing region. In particular, optDMD models extreme values in the spectrogram more accurately in the training region, while exact DMD shows evident underestimation of the PSD magnitudes as well as temporal decay. We believe this is because optDMD optimizes the DMD mode amplitudes based on all of the training data, not just the first snapshot and we enforce limits on the real part of the optDMD eigenvalues. However, optDMD still has trouble modeling the peaks of the spectral behavior in the testing region, shown by the increased error in the fifth column of Figure \ref{fig:spectrogram_grid_r_14}. The optDMD error plots in the testing region are of similar magnitude to those of exact DMD (fourth column of Figure \ref{fig:spectrogram_grid_r_14}), suggesting both exact DMD and optDMD are unable to capture the full spectro-temporal behavior of the state variables in the testing region. Overall, optDMD outperforms exact DMD in recreating the spectrograms of the state variables in the training region, but both algorithms have limitations in forecasting spectral behavior in the testing region. We note that there are additional adjustments we can make to the optDMD code, such as adjusting the initial guess of the eigenvalues or optimization parameters, that could improve its accuracy, especially in the testing region. 

\begin{figure}
    \centering
    \includegraphics[width=1\linewidth]{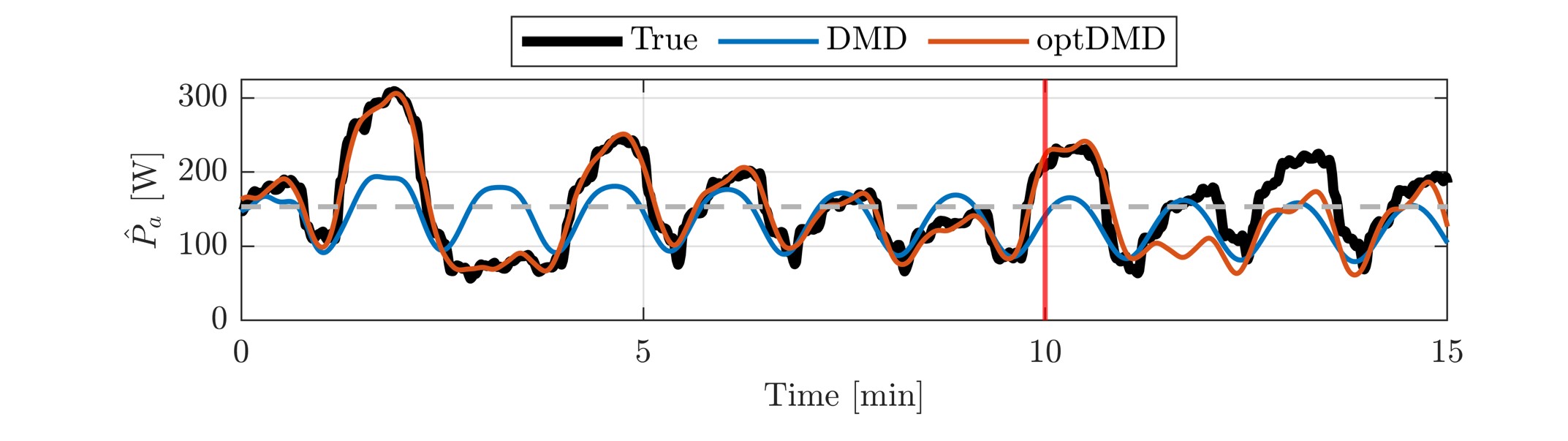}
    \caption{Time series plot of average absorbed power, $\hat{P}_{a}(t)$, with true value in black, the DMD fit in blue, and optDMD in orange. Red line separates training region (left) and testing region (right), and the gray dashed line represents the true average absorbed power over the full time region, calculated from WEC-Sim.}
    \label{fig:absorbed_power_r_14}
\end{figure}

To further evaluate the performance of DMD and optDMD, we consider how well these algorithms can model the amount of power the OSWEC generates over time, or absorbed power. Since the device is operating in irregular seas, the power absorbed by the device can vary significantly in time, making electricity generation and storage difficult. This value is important to know and predict because it can inform when the device is generating more power than necessary and needs to shed power to accommodate energy storage constraints, or when energy generation is low and the device can operate at full efficiency. Because we cannot model the actual velocity time series to calculate instantaneous power absorption, we instead look at the average power absorbed by the OSWEC over all frequencies, $\hat{\boldsymbol{P}}_{a}(t)$. Because our OSWEC model uses a linear PTO, the average power absorbed over a time period, $T_{w}$ is $\hat{\boldsymbol{P}}_{a}(t) = \frac{\nu_{PTO}}{T_{w}} \int_{t}^{t+T_{w}} \left| \dot{\boldsymbol{\theta}}(\tau) \right|^{2} d\tau$. By Parseval's theorem, we can approximate $\hat{\boldsymbol{P}}_{a}(t)$ by taking each ``slice'' of the $\dot{\boldsymbol{\theta}}$ spectrogram in time and integrating the power spectral density over all frequencies, as:
\begin{equation}
    \hat{\boldsymbol{P}}_{a}(t) = \nu_{PTO} \int_{f} \boldsymbol{S}_{\dot{\boldsymbol{\theta}}}(f,t) df.
\end{equation}

Figure \ref{fig:absorbed_power_r_14} shows $\hat{\boldsymbol{P}}_{a}(t)$ calculated from the true spectrograms generated from WEC-Sim data as well as both exact DMD and optDMD spectrograms from Figure \ref{fig:spectrogram_grid_r_14}. Similar to the previous results, optDMD outperforms exact DMD in modeling the average absorbed power, especially in the training region where optDMD models $\hat{\boldsymbol{P}}_{a}(t)$ with near perfect accuracy. The DMD estimate of $\hat{\boldsymbol{P}}_{a}$ oscillates at about the same frequency as the true value, but does not accurately model its magnitude. This suggests that DMD can only identify the low rank dynamics of $\boldsymbol{S}_{\dot{\boldsymbol{\theta}}}$, which is also evident in Figure~\ref{fig:spectrogram_grid_r_14}. While optDMD matches the true values early on in the testing region, neither algorithm accurately forecasts the time evolution of absorbed power consistently in the training region. This again highlights the benefits of using optDMD to model training data and inspires future work to improve accuracy in its ability to forecast future state behavior in a time-resolved manner.

This analysis shows that for an irregular wave field with significant underlying structure, both DMD and optDMD can model the temporal behavior of the power spectrum of the state variables with appropriate rank and training time. However, optDMD is significantly more accurate in the training region and, unlike for a regular wave, neither algorithm performs as well in the testing region. This suggests that these methods could be helpful for state identification in irregular seas, but are less helpful for forecasting. In real seas, we would expect to see additional phenomena such as frequency drifting and nonlinear wave interactions, which could affect the accuracy of DMD and optDMD in the training region. This gap highlights the need for further research using measurements of WEC performance in irregular wave fields.

\section{Conclusion}
\label{conclusion}
This study demonstrates the accuracy with which dynamic mode decomposition can model and predict OSWEC behavior using only physically attainable measurement data from the system dynamics without knowledge of the past or future incident wave field. Specifically, we tested the ability for DMD to model OSWEC behavior when considering three common WEC modeling challenges: noisy signals, nonlinear behavior, and irregular wave input. We showed that for noisy data, exact DMD breaks down with relatively low levels of noise (SNR < 50), but total-least-squares DMD can significantly improve the system model in both the training and testing region with minimal modifications to the original algorithm. Second, we showed that although DMD is a linear algorithm, it can model weakly nonlinear dynamics with reasonable accuracy. When nonlinear dynamics are present, adding even a single time delay can significantly improve DMD accuracy with the same amount of training time by increasing the available rank of the data matrices. Finally, we showed we can describe OSWEC behavior in response to irregular waves by transforming the state variables to the frequency domain and modeling spectro-temporal behavior of the system states. We showed that optimized DMD can significantly improve accuracy in recreating the spectrograms of system states, particularly in the training region. Analysis of the irregular wave case also highlights the limitations of a predictive model without knowledge of the future, stochastic wave field. 

Overall, we see potential for data-driven algorithms, like DMD, to bridge the gap between the wave energy and data science fields and solve common problems in WEC modeling and control. In this study, we highlight the capabilities of DMD to accurately model and predict WEC dynamics in non-ideal conditions, but acknowledge that data from experiments and field deployments will include combinations of the cases presented in this paper along with other complexities not considered. Consequently, future work should apply these methods to experimental or field systems to advance the ultimate objective of using data-driven models to inform model predictive control schemes and optimize power absorption of these devices. With these applications in mind, DMD is a promising tool that has significant potential to support the development of the wave energy field.

\section*{Acknowledgments}
The authors acknowledge support from the National Science Foundation AI Institute in Dynamic Systems (grant number 2112085), Department of Defense Naval Facilities Engineering Systems Command (contract N0002421D6400/N0002421F8712), and the National Science Foundation Graduate Research Fellowship (grant number DGE-2140004).

\begin{appendices}

\section{Dynamic mode decomposition}
\label{sec:dmdappendix}
This appendix includes a detailed derivation of the exact dynamic mode decomposition (DMD) algorithm, as defined in Tu et al.~\cite{tu2013dynamic}. This algorithm is also described in detail in~\cite{kutz2016dynamic}, which includes sample code and data.

Consider a generic linear dynamical system with states $\boldsymbol{x}(t)$:
\begin{equation}
    \frac{d\boldsymbol{x}}{dt} = \boldsymbol{\mathcal{A}} \boldsymbol{x},
\end{equation}
where $\mathcal{A}$ describes the system dynamics in continuous time. To model this system in discrete time, we use the linear operator $\boldsymbol{A} = \text{exp}(\boldsymbol{\mathcal{A}} \Delta t)$ to describe the linear dynamics that shift the system forward from time $t_{k}$ to $t_{k+1}$, with $t_{k} = k \Delta t$:
\begin{equation}
    \label{eqn:vectorsappendix}
    \boldsymbol{x}_{k+1} = \boldsymbol{A} \boldsymbol{x}_{k},
\end{equation}
 where $\boldsymbol{x}_{k}$ is a column vector of all state measurements at time $t_{k}$. We can model Equation \ref{eqn:vectorsappendix} using snapshot data at discrete times arranged into data matrices $\boldsymbol{X}$ and $\boldsymbol{X}^{'}$, such that:
\begin{equation}
    \boldsymbol{X}^{'} \approx \boldsymbol{AX},
    \label{eqn:regressionappendix}
\end{equation}
where the data matrices are of the form:
\begin{equation}
\boldsymbol{X} = 
\begin{bmatrix}
| & | & & | \\
\boldsymbol{x}_{1} & \boldsymbol{x}_{2} & ... & \boldsymbol{x}_{m} \\
| & | & & | \\
\end{bmatrix}, \hspace{0.3cm} \boldsymbol{X}^{'} = 
\begin{bmatrix}
| & | & & | \\
\boldsymbol{x}_{2} & \boldsymbol{x}_{3} & ... & \boldsymbol{x}_{m+1} \\
| & | & & | \\
\end{bmatrix} 
\in \mathbb{R}^{n x m}.
\end{equation}

The solution to the least-squares regression problem laid out in Equation \ref{eqn:regressionappendix} can be approximated by finding a matrix $\boldsymbol{A}$ that minimizes $||\boldsymbol{X}^{'} - \boldsymbol{AX}||_{F}$, where $|| \cdot ||_{F}$ is the Frobenius norm. This can be done using the Moore-Penrose pseudoinverse of data matrix $\boldsymbol{X}$, denoted  $\boldsymbol{X\dag}$:
\begin{equation}
    \boldsymbol{A} = \boldsymbol{X}^{'} \boldsymbol{X}^{\dag}.
\end{equation}
 We calculate $\boldsymbol{X}^{\dag}$ using the singular value decomposition (SVD) of data matrix $\boldsymbol{X}$:
\begin{equation}
    \boldsymbol{X} = \boldsymbol{U \Sigma V}^{*},
\end{equation}
 where the columns of $\boldsymbol{U} \in \mathbb{R}^{n x n}$ and $\boldsymbol{V} \in \mathbb{R}^{m x n}$ are the left and right singular vectors of $\boldsymbol{X}$, respectively, and the diagonal of $\boldsymbol{\Sigma} \in \mathbb{R}^{n x n}$ contains the associated singular values. For large systems, it may be necessary to truncate these matrices and only take the first $r$ singular values and modes to approximate a high-dimensional data matrix, such that:
\begin{equation}
    \boldsymbol{X} \approx \boldsymbol{U}_{r} \boldsymbol{\Sigma}_{r} \boldsymbol{V}^*_{r},
\end{equation}
 where $\boldsymbol{U}_{r} \in \mathbb{R}^{n x r}$, $\boldsymbol{\Sigma}_{r} \in \mathbb{R}^{r x r}$, and $\boldsymbol{V}_{r} \in \mathbb{R}^{m x r}$. Because $\boldsymbol{X}$ is real,  $\boldsymbol{U}_{r}$ and $\boldsymbol{V}_{r}$ are orthogonal, which means the psuedoinverse is easily calculated as $\boldsymbol{X}^{\dag} = \boldsymbol{V}_{r} \boldsymbol{\Sigma}^{-1}_{r} \boldsymbol{U}^{*}_{r}$:
\begin{equation}
    \boldsymbol{A} = \boldsymbol{X}^{'} \boldsymbol{X}^{\dag} = \boldsymbol{X}^{'} \boldsymbol{V}_{r} \boldsymbol{\Sigma}^{-1}_{r} \boldsymbol{U}^{*}_{r} \in \mathbb{R}^{n x n}.
\end{equation}
The eigenvectors of $\boldsymbol{A}$ are the DMD modes, $\boldsymbol{\phi}_{j}$, with corresponding eigenvalues $\lambda_j$ that represent their growth/decay rates and frequencies. For large systems, working with (n x n) matrix $\boldsymbol{A}$ is too computationally expensive. To ease the computational burden, we project $\boldsymbol{A}$ onto the POD modes described by $\boldsymbol{U}_{r}$:
\begin{equation}
    \Tilde{\boldsymbol{A}} \doteq \boldsymbol{U}^{*}_{r} \boldsymbol{A} \boldsymbol{U}_{r} = \boldsymbol{U}^{*}_{r} \boldsymbol{X}^{'} \boldsymbol{V}_{r} \boldsymbol{\Sigma}^{-1}_{r} \in \mathbb{R}^{r x r}.
\end{equation}

To approximate the DMD modes in the POD basis, we set up an eigenvalue problem with $\Tilde{\boldsymbol{A}}$:
\begin{equation}
    \Tilde{\boldsymbol{A}} \boldsymbol{W} = \boldsymbol{W} \boldsymbol{\Lambda},
\end{equation}
 where the columns of $\boldsymbol{W}$ are the eigenvectors of $\Tilde{\boldsymbol{A}}$, and $\boldsymbol{\Lambda}$ contains the eigenvalues of both $\Tilde{\boldsymbol{A}}$ and $\boldsymbol{A}$. As described by~\cite{tu2013dynamic}, the exact DMD modes are the columns of $\boldsymbol{\Phi}$, where:
\begin{equation}
    \boldsymbol{\Phi} = \boldsymbol{X}^{'} \boldsymbol{V}_{r} \boldsymbol{\Sigma}^{-1}_{r} \boldsymbol{W}.
\end{equation}

With the DMD eigenvectors and eigenvalues, we can now describe the system dynamics:
\begin{equation}
    \boldsymbol{x}(t) \approx \sum_{j=1}^{r} \boldsymbol{\phi_{j}} \text{exp}(\omega_{j} t) b_{j} = \boldsymbol{\Phi} \text{exp}(\boldsymbol{\Omega} t) \boldsymbol{b},
    \label{eqn:dynamicsappendix}
\end{equation}
 where $\omega_{j} = {\text{ln}(\lambda_{j})}/{\Delta t}$ are the continuous-time DMD eigenvalues and $\boldsymbol{b} = \boldsymbol{\Phi}^{\dag} \boldsymbol{x}_{1}$ is the initial time vector translated to the DMD eigenbasis. With Equation \ref{eqn:dynamicsappendix}, we can now describe and predict system state dynamics using only snapshot data.

 \subsection{Total-least-squares DMD}
 \label{sec:TLSDMDappendix}
This appendix includes details of the total-least-squares DMD algorithm introduced in Section \ref{sec:TLSDMD} and used in Section \ref{sec:noise}. We refer the reader to \cite{hemati2017biasing} for a full derivation, and to~\cite{kutz2016dynamic} for example code and data.
 
 Consider a new combined data matrix $\boldsymbol{Z}$ and its SVD:
\begin{equation}
    \boldsymbol{Z} = 
    \begin{bmatrix}
        \boldsymbol{X} \\
        \boldsymbol{X}^{'}
    \end{bmatrix}
    = \boldsymbol{U}_{Z} \boldsymbol{\Sigma}_{Z} \boldsymbol{V}^{*}_{Z}.
\end{equation}
 To account for noise in both $\boldsymbol{X}$ and $\boldsymbol{X}^{'}$, we project the data matrices onto a basis created by the right singular vectors of $\boldsymbol{Z}$, $\boldsymbol{V}_{Z}$:
\begin{equation}
    \Breve{\boldsymbol{X}} = \boldsymbol{X} \boldsymbol{V}_{Z} \boldsymbol{V}^{*}_{Z}, \hspace{0.5cm} \Breve{\boldsymbol{X}^{'}} = \boldsymbol{X}^{'} \boldsymbol{V}_{Z} \boldsymbol{V}^{*}_{Z}.
\end{equation}

We now have a modified regression problem: 
\begin{equation}
    \Breve{\boldsymbol{X}^{'}} = \Breve{\boldsymbol{A}} \Breve{\boldsymbol{X}},
\end{equation}
 and can calculate the DMD matrix $\Breve{\boldsymbol{A}}$ using the SVD of $\Breve{\boldsymbol{X}}$:
\begin{equation}
    \Breve{\boldsymbol{A}} = \Breve{\boldsymbol{U}^{*}}  \Breve{\boldsymbol{X}^{'}}  \Breve{\boldsymbol{V}}  \Breve{\boldsymbol{\Sigma}}^{-1},
\end{equation}
 where $\Breve{\boldsymbol{U}}$ and $\Breve{\boldsymbol{V}}$ and right singular vectors of $\Breve{\boldsymbol{X}}$, respectively,  and $\Breve{\boldsymbol{\Sigma}}$ contains the singular values of  $\Breve{\boldsymbol{X}}$, such that $\Breve{\boldsymbol{X}} = \Breve{\boldsymbol{U}} \Breve{\boldsymbol{\Sigma}} \Breve{\boldsymbol{V}}^{*}$. We can now follow the original DMD algorithm, replacing the original linear operator $\boldsymbol{A}$ with $\Breve{\boldsymbol{A}}$.

 \subsection{Time delays}
 \label{sec:TDappendix}
This appendix explains how to modify the data matrices using time delays to increase the rank of the data matrices and gain important phase information \cite{tu2013dynamic}. We introduce using time delays in Section \ref{sec:timedelays} and utilize them in Section \ref{sec:nonlinear}.
 
 By stacking matrices that are shifted by a time step, we obtain new data matrices: 
\begin{equation}
    \boldsymbol{Y} = 
    \begin{bmatrix}
        \boldsymbol{X} \\
        \boldsymbol{X}^{'}
    \end{bmatrix},  \hspace{0.3cm} \boldsymbol{Y}^{'} = 
    \begin{bmatrix}
        \boldsymbol{X}^{'} \\
        \boldsymbol{X}^{''}
    \end{bmatrix},
\end{equation}
where $\boldsymbol{X}^{''}$ follows the same pattern as $\boldsymbol{X}^{'}$, but is shifted two time steps rather than one. $\boldsymbol{Y}$ and $\boldsymbol{Y}^{'}$ replace $\boldsymbol{X}$ and $\boldsymbol{X}^{'}$ in the exact DMD algorithm in Section \ref{sec:dmd}, respectively. Since the maximum rank we can consider when reconstructing $\boldsymbol{x}(t)$ is the length of the column vectors in the data matrices, time delays are particularly useful when higher rank is required for fidelity, but the number of sensors is limited. For example, a single time delay doubles the length of the column vectors in the data matrices, which doubles the number of SVD modes we can use to model the dynamics. Adding a shifted time matrix also provides more phase information, which can improve reconstruction accuracy, especially for the case of standing waves~\cite{tu2013dynamic}.

\subsection{Optimized DMD}
\label{sec:optDMDappendix}
This appendix provides a brief derivation of the main ideas of optDMD that is introduced in Section \ref{sec:optDMD} and used in Section \ref{sec:irregular}. For a complete derivation as well as open-source code, we refer the reader to~\cite{askham2018variable}. 

To begin, we assume our data $\boldsymbol{X}$ is a solution of a linear system of $r$ differential equations, where $r$ is the chosen rank, which means we can express $\boldsymbol{X}$ as a combination of exponentials,
\begin{equation}
    \boldsymbol{X}^{T} \approx \boldsymbol{\Lambda}(\boldsymbol{\alpha})\boldsymbol{B},
\end{equation}
 where $\boldsymbol{\Lambda} \in \mathbb{C}^{mxr}$ is a matrix of exponentials, such that $\Lambda(\boldsymbol{\alpha})_{i,j} = e^{\alpha_{j} t_{i}}$, $\boldsymbol{\alpha}$ are the equivalent of continuous-time DMD eigenvalues, and $\boldsymbol{B} \in \mathbb{C}^{rxn}$ are the weights of the exponential functions. To find the optimal values of $\boldsymbol{\alpha}$ and $\boldsymbol{B}$ that best approximates our data matrix $\boldsymbol{X}$, we want to solve the following nonlinear least squares problem:
\begin{equation}
    \min_{\boldsymbol{\alpha}, \hspace{0.07cm} \boldsymbol{B}} \left| \left| \hspace{0.03cm}
    \boldsymbol{X}^{T} - \boldsymbol{\Lambda}(\boldsymbol{\alpha})\boldsymbol{B}
    \hspace{0.03cm} \right| \right|_{F},
    \label{eqn:nonlinear_least_squares}
\end{equation}
 which is achieved using variable projection methods outlined in~\cite{askham2018variable}. In Eq.~\ref{eqn:nonlinear_least_squares}, $F$ denotes the Frobenius norm. Once we have $\hat{\boldsymbol{\alpha}}$ and $\hat{\boldsymbol{B}}$ that minimizes Equation \ref{eqn:nonlinear_least_squares}, we can calculate the DMD eigenvalues and eigenvectors:
\begin{equation}
    \gamma_{i} = \hat{\alpha}_{i}, \hspace{0.3cm} \boldsymbol{\phi}_{i} = \frac{1}{\left| \left| \hat{\boldsymbol{b}}_{i} \right| \right|_{2}} \hat{\boldsymbol{b}}_{i},
\end{equation}
 and estimate system dynamics:
\begin{equation}
    \boldsymbol{x}(t) = \sum_{i=1}^{r} \left| \left| \hat{\boldsymbol{b}}_{i} \right| \right|_{2} e^{\gamma_{i} t} \boldsymbol{\phi}_{i},
\end{equation}
where $\hat{\boldsymbol{b}}_{i}$ is the ith column of $\hat{\boldsymbol{B}}^{T}$.

\section{Semi-analytical model and WEC-Sim comparison}
\label{sec:diaswecsimcompare}
\begin{figure}[t!]
    \centering
    \includegraphics[width=3in]{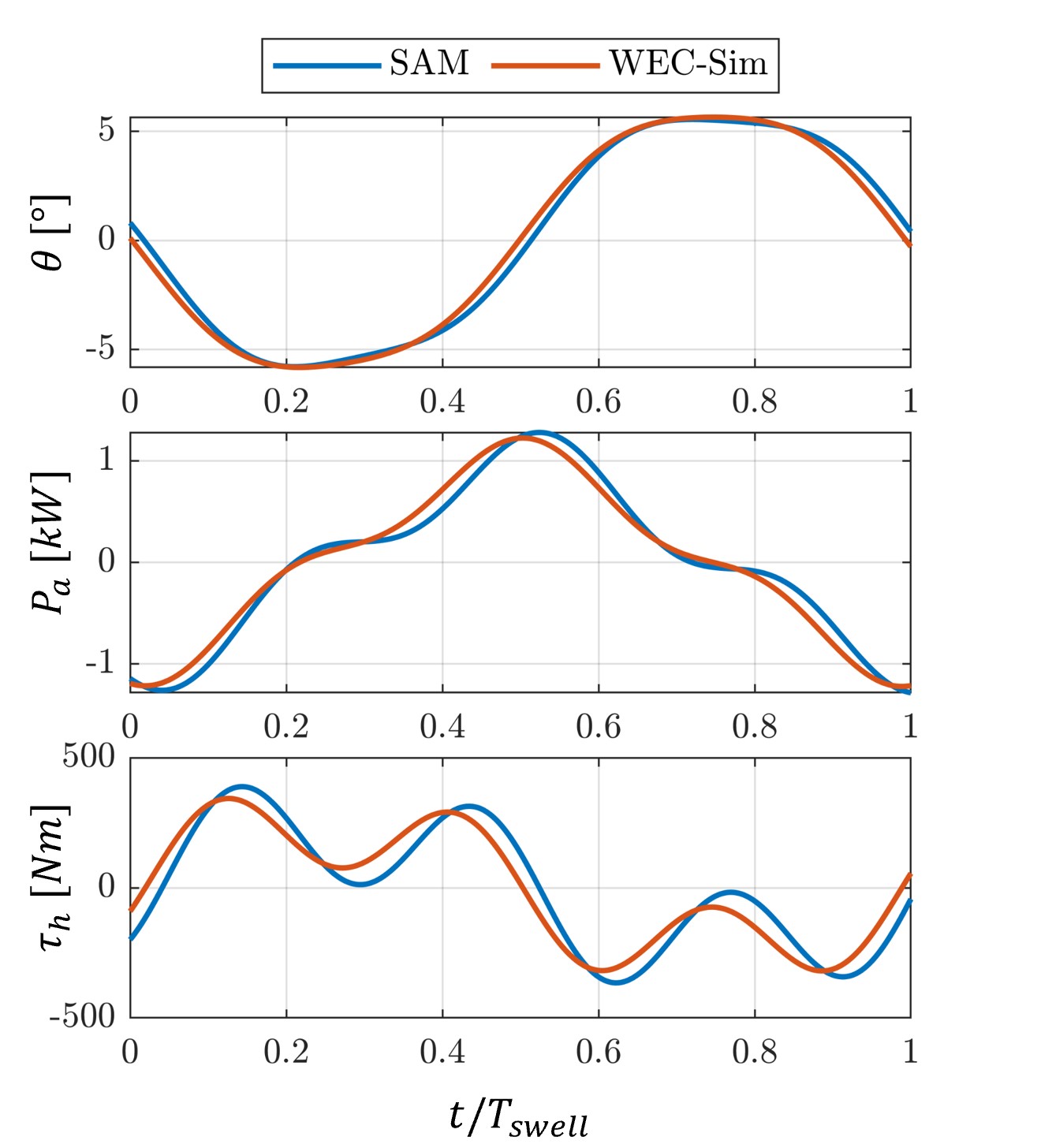}
    \caption{Time series plots of angular rotation, $\boldsymbol{\theta}$ (top), absorbed power, $\boldsymbol{P}_{a}$ (middle), and hydrodynamic torque, $\boldsymbol{\tau}_{h}$ (bottom), calculated using the semi-analytical model (SAM) from Renzi et al. \cite{renzi2013hydrodynamics} (blue) and WEC-Sim (orange).}
    \label{fig:diaswecsimcomparefigure}
\end{figure}

In this section, we present a comparison between the semi-analytical model from Renzi et al.~\cite{renzi2013hydrodynamics} used in Case 1, and the open-source code WEC-Sim \cite{wecsim} used in Cases 2 and 3. We consider the outputs of each method in response the the incident wave used in Case 1 (Section \ref{sec:noise}). As summarized in Table \ref{tab:dmd_cases}, the wave input is a sum of two monochromatic waves, with wave periods of 8 and 2.55 seconds and corresponding wave heights of 0.75 and 0.3 meters, respectively. We compare three parameters: angular rotation, $\boldsymbol{\theta}$, power absorbed by the device, $\boldsymbol{P}_{a} = \nu_{PTO} \dot{\boldsymbol{\theta}}$, and hydrodynamic torque, $\boldsymbol{\tau}_{h}$. For the angular rotation results from the semi-analytical model, we used a amplitude scale of $A_{i} = w/25 = 0.72$. This value is not formally defined in \cite{renzi2013hydrodynamics}, but just described as being ``much smaller'' than the flap width. We found this expression to fit the WEC-Sim kinematic data well over multiple wave amplitudes. The results of the comparison are shown in Figure \ref{fig:diaswecsimcomparefigure}.

We see excellent agreement between the two modeling methods in both amplitude and phase. We note that the amplitude scale, $A_{i}$, affects the agreement in amplitude for the kinematics, but the agreement in phase and relative magnitude of the two wave components implies there is still an underlying agreement of state behavior independent of the amplitude scale fitting. Consequently, we are confident that both the semi-analytical model and WEC-Sim represent similar dynamics in the linear regime.

\end{appendices}

\newpage
\bibliography{references}
\bibliographystyle{unsrt}

\end{document}